\documentclass[prd,twocolumn,superscriptaddress,showpacs]{revtex4}
\usepackage{latexsym}
\usepackage{amsmath}
\usepackage{amssymb}
\usepackage[dvips]{graphicx}
\usepackage{color}
\newcommand{\BM}[1]{\textcolor{black}{ #1 }}
\newcommand{\JFD}[1]{\textcolor{black}{ #1 }}
\newcommand{\rd}{\mathrm{d}}
\newcommand{\calC}{\mathcal{C}}

\newcommand{\calN}{\mathcal{N}}

\newcommand{\calW}{\mathcal{W}}

\newcommand{\Slash}[1]{#1\!\!\!/}
\newcommand{\kslash}{k\!\!\!/}

\newcommand{\pslash}{\partial \kern -0.55em/ }
\newcommand{\calCstar}{\hat{\mathcal{C}}}

\newcommand{\Comm}[2]{\ensuremath{[ \, #1, #2 \, ]}}
\newcommand{\AntiComm}[2]{\ensuremath{\{ \, #1, #2 \, \}}}

\newcommand{\Tr}[1]{\mathrm{Tr} [ \, #1 \, ] }

\newcommand{\expect}[1]%
   {\ensuremath{\langle \, #1 \, \rangle}}
\newcommand{\expectTproduct}[1]%
   {\ensuremath{\langle \, \mathcal{T} \{ \, #1 \, \} \, \rangle}}
\begin{document}
%
%
%
\preprint{LAUR LA-UR-05-8819}

\title{Supersymmetric approximations to the 3D supersymmetric $O(N)$ model}
\author{John F. Dawson}
\email{john.dawson@unh.edu}
\affiliation{Department of Physics,
   University of New Hampshire, Durham, NH 03824}
\author{Bogdan Mihaila}
\email{bogdan.mihaila@unh.edu} 
\affiliation{Theoretical Division,
   Los Alamos National Laboratory, Los Alamos, NM 87545}
\author{Per Berglund}
\email{per.berglund@unh.edu}
\affiliation{Department of Physics,
   University of New Hampshire, Durham, NH 03824}
\author{Fred Cooper}
\email{cooper@santafe.edu}
\affiliation{Theoretical Division,
   Los Alamos National Laboratory, Los Alamos, NM 87545}
\affiliation{Santa Fe Institute,
   1399 Hyde Park Road, Santa Fe, NM 87501}
%
%
\begin{abstract}
We develop several non-perturbative approximations for studying the
dynamics of a supersymmetric $O(N)$ model which preserve
supersymmetry. We study the phase structure of the vacuum in both the
leading order in large-$N$ approximation as well as in the Hartree
approximation, and derive the finite temperature renormalized
effective potential. We derive the exact Schwinger-Dyson equations for
the superfield Green functions and develop the machinery for going
beyond the next to leading order in large-$N$ approximation using a
truncation of these equations which can also be derived from a
two-particle irreducible effective action.
\end{abstract}
\maketitle
%
%
%
\section{Introduction}
\label{s:introduction}

Theories with supersymmetry (SUSY) have been very attractive to
theoretical physicists because they solve the problem of taming the
quadratic divergences associated with mass renormalization of scalar
fields~\cite{ref:WessBagger92}. This cancellation of mass corrections
when one includes the related boson and fermion loops is most apparent
in the superspace formulation of supersymmetry.  If supersymmetry
turns out to be a good representation of reality, it would be nice to
have approximate analytical methods of understanding the phase
structure and dynamics of these theories.  Recent advances in
approximation schemes to field theory have shown that approximations
based on two-particle irreducible (2-PI) effective
actions~\cite{r:CJT,r:LW,r:Baym62} have the potential of leading to
thermalization of quantum
fields~\cite{r:BC01,r:AB01,r:B02,r:AABBS,r:CDM02ii,r:CDM02}. These
approximations also allow one to study the dynamics of phase
transitions when the appropriate order parameter is found.  What we
would like to show here is that the methodology used in obtaining the
aforementioned approximations in scalar field theories, can easily be
generalized to the supersymmetric extension of the theory. In fact,
when the superspace formalism is in terms of polynomial interactions
of scalar superfields, standard field theory approximations such as
large-$N$ expansions \cite{r:W73,r:CJT,r:CJP}, Hartree approximations
\cite{r:S74,r:C75,r:CPS86,r:CMprd36,r:PS86,r:BVH94} and their
resummations via self consistent Schwinger-Dyson equation methods (or
effective 2-PI actions) automatically preserve supersymmetry at zero
temperature.

The new feature in this work that differentiates it from our previous
studies of scalar $\phi^4$ field
theory~\cite{r:CDM02ii,r:CDM02,r:BCDM01,r:MCD01} is that the
superfields now depend on anticommuting Grassmann variables as well as
the usual space-time coordinates and the action includes integration
not only over Minkowski space (here 2+1 dimensional) but also over the
two component Majorana spinor $\theta$ of Grassmann coordinates. The
superfields contain both bosonic and fermionic degrees of freedom with
the interactions dictated by the need for invariance under the
supersymmetry transformations.

At finite temperature, supersymmetry is softly
broken~\cite{r:DasKaku78}.  However this occurs in a way which does not
affect the cancellation of ultraviolet divergences, since the finite
temperature modifications of the super-propagators only affects the
infrared physics. Thus the use of supergraphs maintains its usefulness
even at finite temperature.

The model we will study is the $O(N)$ supersymmetric $\Phi^4$ model,
which is actually a scalar $\phi^6$ field theory interacting with
fermions in a manner consistent with SUSY.  This model has recently
been studied by Moshe and Zinn-Justin~\cite{ref:MMZJ} (referred to as
MZJ in what follows) and by Feinberg, Moshe, and
Smolkin~\cite{ref:FMS} in 2+1 dimensions at finite temperature in the
leading order in large-$N$ approximation.  Their interest was mainly
in the spontaneous breakdown of scale invariance but they also found
an interesting phase structure which depended on the sign of the
renormalized mass parameter as well as the value of the renormalized
coupling constant.  In this work we will formulate the same model in a
slightly more convenient way using the Hubbard-Stratonovich formalism.
We will compare the leading order in large-$N$ approximation to the
Hartree approximation. We will find that although the two
approximations lead to identical dynamics when the expectation value
of $\phi$ is zero, the ground states found in these two approximations
are quite different and lend themselves to exploring different types
of phase transitions.  In both approximations the vacuum is
degenerate.  For some choices of the parameters on finds in both
approximations that the states with zero and non zero expectation
value of $\phi$ can coexist.  This possibility leads to interesting
dynamical questions of how an initial state prepared at high
temperature and then allowed to expand would choose one or the other
vacuum. In this paper we also derive the exact Schwinger-Dyson
equations for the superfields in terms of the auxiliary fields with a
future goal of doing dynamical simulations as well as studying whether
the vacuum degeneracy gets lifted.

In what follows we will use as much as possible the notation of
earlier studies of the phase structure of these models which is found
in the work of Moshe and Zinn-Justin~\cite{ref:MMZJ} and Shifman,
Vainshtein, and Voloshin~\cite{ref:ShifVainVolo99}.  The paper is
organized as follows.  In section II we discuss the minimal
supersymmetric action and derive the large-$N$ and Hartree
approximations. We also derive the exact Schwinger-Dyson equations and
derive two related approximations that resum the next to leading order
large-$N$ approximation. In section three we derive the effective
potential for both the leading order large-$N$ and Hartree
approximations and discuss the phase structure of the vacuum as well
as the behavior of the effective potential at finite temperature. We
summarize our results in section IV.

%
%
\section{Minimal supersymmetric action}
\label{s:action}

The minimal action for $N$ commuting superfields in $d$ space-time
dimensions for $2 \le d \le 3$ is given by:
\begin{equation}\label{e:superactionI}
\begin{split}
   S[\Phi]
   &=
   \int \rd s \>
   \biggl \{ \,
   \frac{1}{2} \,
   \bigl [ \bar{D} \, \Phi_i(s) \bigr ] \cdot
   \bigl [ D \, \Phi_i(s) \bigr ]
   \\  & \qquad\qquad\qquad
   +
   2 N \, \calW \bigl [ \Phi(s)/\sqrt{N} \bigr ] \,
   \biggr \} \>,
\end{split}
\end{equation}
where $s = (x^{\mu},\theta_a)$ with $\mu = 0,\dotsc,d-1$ and $a = 1,2$
and where we have used a summation convention for the $N$ superfields
with $i=1,\dotsc,N$.  The integration measure $\rd s$ is given by:
\begin{equation}
   \rd s
   =
   \rd^d x \, \rd^2\theta
   =
   \rd^d x \, \rd \bar{\theta}_1\rd \theta_1 / 2
   =
   i \>  \rd^d x \, \rd\theta_2 \, \rd\theta_1 \>.
\end{equation}

The superfields $\Phi_i(s)$ can be expanded into commuting and
anticommuting components.  We write:
\begin{equation}\label{e:superfieldsdef}
   \Phi_i(s)
   =
   \phi_i(x)
   +
   \bar\theta \cdot \psi_i(x)
   +
   \frac{1}{2} \,
   \bar\theta \cdot \theta \, F_i(x) \>.
\end{equation}
The superfields \emph{commute} at the same superspace point.
Superderivative spinors are defined by:
\begin{equation}\label{e:superderivatives}
   D
   =
   +
   \bar{\partial}
   -
   i \, \Slash{\partial} \cdot \theta \>,
   \qquad
   \bar{D}
   =
   -
   \partial
   +
   i \, \bar{\theta} \cdot  \Slash{\partial} \>,
\end{equation}
where $\partial$ and $\bar{\partial}$ are Grassmann derivatives with
respect to $\theta$ and $\bar{\theta}$ respectively.  Properties of
the superderivatives are further discussed in
appendix~\ref{a.s:grassmann}.

\begin{figure}[b]
   \centering
   \includegraphics[width=3.0in]{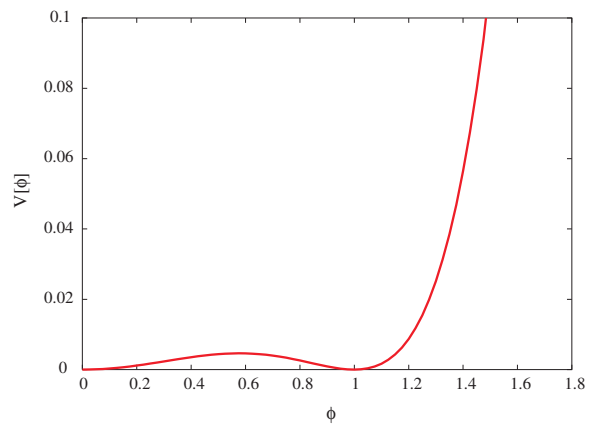}
   \caption{(Color online) Plot of the classical potential for $N=1$ as a function of
   $\phi$ for the case $\lambda =1$, $\phi_0 = 1$.}
   \label{f:Vclassical}
\end{figure}

For the $O(N)$ model, we choose a superpotential of the form:
\begin{equation}\label{e:Wsuper}
   2 N \, \calW \bigl [ \Phi(s)/\sqrt{N} \bigr ]
   =
   \frac{\lambda}{8 N} \,
   \Bigl [ \,
      \Phi_i^2(s)
      -
      N \, \phi_0^2 \,
   \Bigr ]^2 \>,
\end{equation}
where $\phi_0$ is a constant (non-Grassmann).  This potential is
\emph{fourth} order in the superfields but \emph{sixth} order in the
scalar fields.  In terms of component fields, the action
\eqref{e:superactionI} becomes:
\begin{multline}\label{e:spacetimeactionI}
   \!\! S[\phi,\psi,F]
   =
   \frac{1}{2} \!
   \int \! \rd^d x \,
   \Bigl \{
      \bigl [ \partial_{\mu} \phi_i(x) \bigr ] \,
      \bigl [ \partial^{\mu} \phi_i(x) \bigr ]
      +
      F_i^2(x) \
      \\
      +
      \frac{\lambda}{N} \, X(x) \, \phi_i(x) F_i(x)
      \\
      +
      \bar{\psi}_i(x) \cdot
      \bigl ( \,
         i \, \delta_{ij} \, \gamma^{\mu} \, \partial_{\mu}
         -
         M_{ij} \,
      \bigr ) \cdot \psi_j(x) \,
   \Bigr \}
\end{multline}
where
\begin{equation}
\begin{split}
   X(x)
   &=
   \frac{1}{2} \
   \Bigl [ \,
      \sum_j \phi_j^2(x) - N \, \phi_0^2 \,
   \Bigr ] \>,
   \\
   M_{ij}(x)
   &=
   \frac{\lambda}{8 N} \,
   \bigl [ \,
      4 \, \delta_{ij} \, X(x)
      +
      \phi_i(x) \, \phi_j(x) \,
   \bigr ]  \>,
\end{split}
\end{equation}
We see here that $F_i(x)$ is not a dynamical variable.  Varying
the action with respect to $F_i(x)$ gives the constraint:
\begin{equation}\label{e:a.Ficonstraint}
   F_i(x)
   =
   - \frac{\lambda}{2 N} \, X(x) \, \phi_i(x) \>.
\end{equation}
Using this result, the action \eqref{e:spacetimeactionI} becomes:
\begin{multline}\label{e:spacetimeactionII}
   S[\phi,\psi]
   =
   \frac{1}{2}
   \int \rd^d x \,
   \Bigl \{
      \bigl [ \partial_{\mu} \phi_i(x) \bigr ] \,
      \bigl [ \partial^{\mu} \phi_i(x) \bigr ]
      -
      2 \, V(\phi)
      \\
      +
      \bar{\psi}_i(x) \cdot
      \bigl ( \,
         i \, \delta_{ij} \, \gamma^{\mu} \, \partial_{\mu}
         -
         M_{ij} \,
      \bigr ) \cdot \psi_j(x) \,
   \Bigr \} \>.
\end{multline}
where
\begin{equation}
   V(\phi)
   =
   \frac{\lambda^2}{8 N^2} \, X^2(x) \,
   \bigl [ \,
      2 \, X(x)
      +
      N \, \phi_0^2 \,
   \bigr ]  \>.
\end{equation}
A graph of the classical scalar potential $V(\phi)$ for $N=1$,
$\lambda=1$, and $\phi_0 = 1$, as a function of $\phi$ is shown in
Fig.~\ref{f:Vclassical}.  The curve is symmetric about the origin,
with two minima's at $\phi = 0$ and $\phi = \phi_0$.

%
%
\subsection{Large-$N$ approximation}

For the large-$N$ approximation, it is easier to count powers of
$1/N$ by introducing a commuting composite superfield $\chi(s)$.  In
general, this can be done for an arbitrary polynomial Lagrangian by
introducing a functional delta function of the type:
\begin{equation}
   \int \rd \chi \,
   \delta
   \Bigl [ \,
      \chi(x)
      -
      \frac{\lambda}{2N} \, \phi_i^2(x) \,
   \Bigr ]
   =
   1 \>,
\end{equation}
into the path integral.  This is the strategy used by MZJ. However,
for quartic scalar interactions, it is simpler to use the
Hubbard-Stratonovich transformation to convert the quartic term into
a Gaussian at the cost of an additional integration, using the
identity:
\begin{multline}
   \int \rd \phi \,
   \exp
   \bigl ( \,
   -
   \tfrac{1}{2} \, \phi \, G^{-1} \, \phi
   +
   j \, \phi \,
   \bigr )
   \\
   =
   \exp
   \bigl \{ \,
      \tfrac{1}{2} \, j \, G \, j
      -
      \tfrac{1}{2} \, \Tr{\ln \, G^{-1} } \,
   \bigr \} \>,
\end{multline}
with $j$ being proportional to $\phi_i^2(x)$.  The same trick
applies to the superfield case.  This is equivalent to introducing a
commuting composite super field $\chi(s)$ by subtracting from the
action \eqref{e:superactionI} a term of the form:
\begin{equation}\label{e:addtoaction}
   \int \rd s \>
   \frac{N}{2 \lambda} \,
   \biggl \{
      \chi(s) \
      \BM{-} \
      \frac{\lambda}{2 N} \,
      \Bigl [ \,
         \sum_{j=1}^N \Phi_j^2(s) - N \, \phi_0^2 \,
      \Bigr ]
   \biggr \}^2  \>.
\end{equation}
This leads to an equivalent action given by:
\begin{align}\label{e:largeNactionI}
   & S[\Phi,\chi;J,K]
   =
   \\ \notag &
   \int \rd s \,
   \biggl \{ \,
      \frac{1}{2} \,
      \Bigl [ \,
      ( \bar{D} \, \Phi_i(s) ) \cdot
      ( D \, \Phi_i(s) ) \
      \BM{+} \
      \chi(s) \, \Phi_i^2(s) \,
      \Bigr ]
   \\ \notag &
      -
      \frac{N}{\lambda}
      \Bigl [
         \frac{\chi^2(s)}{2}
         \JFD{-}
         \JFD{2 \mu} \, \chi(s)
      \Bigr ]
      +
      J_i(s) \Phi_i(s)
      +
      K(s) \chi(s)
   \biggr \} \>,
\end{align}
where $\JFD{2 \mu = -\lambda \, \phi_0^2 / 2}$ ($\mu$ has units of
mass).
One of the things we will show is that the action
\eqref{e:largeNactionI} reproduces the results of MZJ and leads to a
simpler formula for the corrections to large-$N$.  The
supergenerating functional $Z[J,K]$ is defined by the path integral:
\begin{align}
   Z[J,K]
   &=
   \exp \bigl \{ i W[J,K] \, \bigr \}
   \label{e:pathintegralI} \\
   &=
   \int \rd \chi \prod_{i} \int \rd \Phi_i \>
      \exp \bigl \{ \, i S[\Phi,\chi;J,K] \, \bigr \} \>.
   \notag
\end{align}
\emph{Average} values of the superfields are obtained by
differentiation of the supergenerating functional:
\begin{equation}\label{e:averagefields}
\begin{split}
   \expect{ \Phi_i(s) }
   =
   \frac{1}{i Z} \, \frac{ \partial Z[J,K] }{ \partial J_i(s) }
   =
   \frac{ \partial W[J,K] }{ \partial J_i(s) } \>,
   \\
   \expect{ \chi(s) }
   =
   \frac{1}{i Z} \, \frac{ \partial Z[J,K] }{ \partial K(s) }
   =
   \frac{ \partial W[J,K] }{ \partial K(s) } \>.
\end{split}
\end{equation}
We introduce an inverse Green function $G^{-1}_{ij}[\chi](s,s)$ by:
\begin{equation}
   G^{-1}_{ij}[\chi](s,s')
   =
   \bigl [ \,
      \bar{D} \cdot D \
      \BM{-} \
      \chi(s) \,
   \bigr ] \, \delta_{ij} \, \delta(s,s') \>,
\end{equation}
so that the Green function $G_{ij}[\chi](s,s')$ satisfies the
superdifferential equation:
\begin{equation}\label{e:desuperG}
   \bigl [ \,
      \bar{D} \cdot D \
      \BM{-} \
      \chi(s) \,
   \bigr ] \, G_{ij}(s,s')
   =
   \delta_{ij} \, \delta(s,s') \>.
\end{equation}
Integrating by parts, the action \eqref{e:largeNactionI} can be
written as:
\begin{align}\label{e:largeNactionII}
   S[\Phi,\chi;& J,K]
   \!=\!
   - \frac{1}{2}
   \int \!\! \rd s \!\! \int \! \rd s' \,
      \Phi_i(s) G^{-1}_{ij}[\chi](s,s') \Phi_j(s')
   \notag \\ &
      -
   \int \rd s \,
   \biggl \{ \,
      \frac{N}{\lambda} \,
      \Bigl [ \,
         \frac{\chi^2(x,\theta)}{2} \
         \JFD{-} \
         \JFD{2 \mu} \, \chi(x,\theta) \,
      \Bigr ]
   \notag \\ &
      +
      J_i(s) \, \Phi_i(s)
      +
      K(s) \, \chi(s) \,
   \biggr \} \>.
\end{align}
The action \eqref{e:largeNactionII} is quadratic in the fields
$\Phi_i(s)$ so we can integrate them out of the generating functional.
This gives:
\begin{equation}\label{e:pathintegralII}
   Z[J,K]
   =
   \calN \int \rd \chi \,
      \exp \bigl \{ \, i S'[\chi;J,K] \, \bigr \} \>,
\end{equation}
where $\calN$ is a constant and:
\begin{multline}\label{e:Sprime}
   S'[\chi;J,K]
   =
   \int \rd s
   \biggl \{ \,
      -
      \frac{N}{\lambda} \,
      \Bigl [ \,
         \frac{\chi^2(s)}{2} \
         \JFD{-} \
         \JFD{2 \mu} \, \chi(s) \,
      \Bigr ]
      \\
      +
      K(s) \, \chi(s)
      +
      \frac{i N}{2} \,
      \ln \{ G^{-1}_{ii}[\chi](s,s) \}
   \biggr \}
   \\
   +
   \frac{1}{2}
   \int \rd s \int \rd s' \,
   J_i(s) \, G_{ij}[\chi](s,s') \, J_j(s) \>.
\end{multline}
The integral over $\chi(s)$ in \eqref{e:pathintegralII} is now done by
the method of steepest descent.  We expand the exponent about a
superfield $\chi_0$:
\begin{multline}\label{e:Spexpansion}
   S'[\chi;J,K]
   =
   S'[\chi_0;J,K]
   \\
   +
   \int \rd s \,
   \bigl [ \, \chi(s) - \chi_0(s) \, \bigr ] \,
   \Bigl [ \,
      \frac{\partial S'[\chi;J,K] }{\partial \chi(s)} \,
   \Bigr ]_{\chi_0}
   \\
   +
   \frac{1}{2} \int \rd s \int \rd s' \,
   \bigl [ \, \chi(s) - \chi_0(s) \, \bigr ] \,
   \bigl [ \, \chi(s') - \chi_0(s') \, \bigr ] \,
   \\ \times
   \Bigl [ \,
      \frac{\partial^2 S'[\chi;J,K] }
           {\partial \chi(s) \, \partial \chi(s')} \,
   \Bigr ]_{\chi_0}
   +
   \dotsb
\end{multline}
where we choose $\chi_0$ such that the linear term vanishes.  This
gives the stationary condition:
\begin{multline}\label{e:supergap}
   \chi_0(s)
   =
   2 \mu \
   \BM{+} \
   \frac{\lambda}{2}
   \Bigl [ \,
      \frac{1}{N} \, \Phi_{i}^2[\chi_0,J](s)
      +
      G_{ii}[\chi_0](s,s) / i \,
   \Bigr ]
   \\
   +
   \frac{\lambda}{N} \, K(s) \>.
\end{multline}
Evaluated at $K(s) = 0$, Eq.~\eqref{e:supergap} is the \emph{supergap}
equation.  Here we have defined $\Phi_{i}[\chi_0,J](s)$, which is a
functional of $\chi_0$ and $J$, as the solution of the integral
equation:
\begin{equation}\label{e:Phi0def}
   \Phi_{i}[\chi_0,J](s)
   =
   \int \rd s' \, G_{ij}[\chi_0](s,s') \, J_j(s') \>.
\end{equation}
We have the remaining action, which is given by:
\begin{equation}
   W[J,K]
   =
   W_0[J,K]
   +
   \frac{i}{2} \, \int \rd s \,
   \ln \{ D^{-1}(s,s) \}
   +
   \dotsb
\end{equation}
where
\begin{multline}\label{e:Wzero}
   W_0[J,K]
   =
   \int \rd s
   \biggl \{ \,
      -
      \frac{N}{\lambda} \,
      \Bigl [ \,
         \frac{\chi_0^2(s)}{2} \
         \JFD{-} \
         \JFD{2 \mu} \, \chi_0(s) \,
      \Bigr ]
      \\
      +
      K(s) \, \chi_0(s)
      +
      \frac{i N}{2} \,
      \ln \bigl [  G^{-1}_{ii}[\chi_0](s,s) \bigr ]
   \biggr \}
   \\
   +
   \frac{1}{2}
   \int \rd s \int \rd s' \,
   J_i(s) \, G_{ij}[\chi_0](s,s') \, J_j(s) \>,
\end{multline}
and where
\begin{equation}\label{e:fullDdef}
\begin{split}
   D^{-1}(s,s')
   &=
   \frac{1}{N}
   \Bigl [ \,
      \frac{\partial^2 S'[\chi;J,K] }
           {\partial \chi(s) \, \partial \chi(s')} \,
   \Bigr ]_{\chi_0}
   \\
   &=
   D^{-1}_0(s,s')
   +
   \Pi_0(s,s') \>,
\end{split}
\end{equation}
with:
\begin{equation}\label{e:Dzerodefs}
   D^{-1}_0(s,s')
   =
   - \frac{1}{\lambda} \, \delta(s,s')  \>,
\end{equation}
and
\begin{multline}\label{e:Pizerpdefs}
   \Pi_0(s,s')
   \\
   =
   \frac{1}{N} \,
   \Phi_i[\chi_0](s) \,
   G_{ij}[\chi_0](s,s') \,
   \Phi_j[\chi_0](s')
   \\
   +
   \frac{1}{2i} \,
   G_{ij}[\chi_0](s,s') \,
   G_{ji}[\chi_0](s',s) \>.
\end{multline}
The vertex function is given by a Legendre transformation:
\begin{multline}
   \Gamma[\Phi,\chi]
   =
   W[J,K]
   \\
   -
   \int \rd s \,
   \Bigl \{ \,
       J_i(s) \, \Phi_{i}(s)
      +
       K(s) \, \chi(s) \,
   \Bigr \} \>,
\end{multline}
where
\begin{equation}\label{e:lrgN.chidef}
\begin{split}
   \Phi_i(s)
   &=
   \frac{\partial W[J,K]}{\partial J_i(s)}
   =
   \Phi_{i}[\chi_0](s) + \frac{1}{N} \Phi_{1\,i}(s) + \cdots
   \\
   \chi(s)
   &=
   \frac{\partial W[J,K]}{\partial K(s)}
   =
   \chi_0(s) + \frac{1}{N} \chi_1(s) + \cdots
\end{split}
\end{equation}
So, to first order in $1/N$, we find the effective action:
\begin{align}\label{lN.e:effectiveaction}
   & \Gamma[\Phi,\chi]
   =
   - \frac{1}{2}
   \int \! \rd s \int \! \rd s' \,
   \Bigl \{
      \Phi_i(s)
      G^{-1}_{ij}[\chi](s,s')
      \Phi_j(s')
   \Bigr \}
   \notag \\ &
   -
   \int \rd s \>
   \Bigl \{ \,
      \frac{N}{\lambda} \,
      \Bigl [ \,
         \frac{\chi^2(s)}{2} \
         \JFD{-} \
         \JFD{2 \mu} \, \chi(s) \,
      \Bigr ]
   \\ \notag & \qquad
      +
      \frac{i N}{2} \,
      \ln \bigl [ G^{-1}_{ii}[\chi](s,s) \bigr ]
      +
      \frac{i}{2} \,
      \ln \bigl [ D^{-1}[\chi](s,s) \bigr ]
   \Bigr \} \>,
\end{align}
which is the classical action plus the trace-log terms.

We list again here the superequations to be solved in first order
large-$N$.  We set the currents to zero, and arrive at the following
equations:
\begin{subequations}\label{e:firstorderlargeN}
\begin{gather}
   \bigl [ \,
      \bar{D} \cdot D \
      \BM{-} \
      \chi(s) \,
   \bigr ] \, \Phi_i(s)
   =
   0 \>,
   \label{e:folNi} \\
   \bigl [ \,
      \bar{D} \cdot D \
      \BM{-} \
      \chi(s) \,
   \bigr ] \, G_{ij}(s,s')
   =
   \delta_{ij} \, \delta(s,s') \>,
   \label{e:folNii} \\
   \chi(s)
   =
   2 \mu \
   \BM{+} \
   \frac{\lambda}{2} \,
   \Bigl [ \,
      \Phi_i^2(s) / N
      +
      G_{ii}(s,s) / i \,
   \Bigr ]  \>.
   \label{e:folNiii}
\end{gather}
\end{subequations}
From Eq.~\eqref{lN.e:effectiveaction}, for $N=1$, the effective
large-N superpotential is given by:
\begin{multline}\label{lN.e:superpotential}
   V_{\text{N}}[\Phi,\chi]
   =
   \int \rd^2 \theta \,
   \Bigl \{ \,
   -
   \frac{1}{2} \, ( \bar{\theta} \cdot \theta ) \, F^2
   \BM{-} \
   \frac{1}{2} \, \chi \, \Phi^2
   \\
   \BM{+}
   \frac{1}{\lambda} \,
   \Bigl [ \,
      \frac{\chi^2}{2} \
      \JFD{-} \
      \JFD{2 \mu} \, \chi \,
   \Bigr ]
   -
   \frac{i}{2} \, \Tr{ \ln \, G^{-1}[\Phi,\chi] } \,
   \Bigr \} \>,
\end{multline}
where the first term comes from the kinetic part of the energy.

%
%
\subsection{Hartree equations}
\label{ss:Hartreeeqs}

In this section, we develop the Hartree equations for this system.  We
start with the action given in Eq.~\eqref{e:superactionI}:
\begin{equation}\label{e:superactionII}
\begin{split}
   S[\Phi]
   &=
   \int \rd s \>
   \biggl \{ \,
   \frac{1}{2} \,
   \bigl [ \bar{D} \, \Phi_i(s) \bigr ] \cdot
   \bigl [ D \, \Phi_i(s) \bigr ]
   \\  & \qquad\qquad\qquad
   +
   \frac{\lambda}{8 N} \,
   \bigl [ \,
      \Phi_j^2(s)
      -
      N \, \phi_0^2 \,
   \bigr ]^2 \,
   \biggr \} \>.
\end{split}
\end{equation}
The equations of motion are given by:
\begin{equation}
   \bigl ( \,
      \bar{D} D \JFD{-} \JFD{2 \mu} \,
   \bigr ) \, \Phi_i(s)
   \BM{-}
   \frac{\lambda}{2N} \,
   \Phi_j^2(s) \, \Phi_i(s)
   =
   J_i(s)  \>,
\end{equation}
where $J_i(s)$ is an external supercurrent, and where we have again
set $\JFD{2 \mu = -\lambda \phi_0^2 / 2}$.  Considering this as an
operator equation and taking expectation values gives the classical
equation:
\begin{equation}
   \bigl ( \bar{D} D \BM{-} 2 \mu \bigr ) \expect{\Phi_i(s)}
   \BM{-}
   \frac{\lambda}{2N}
   \expect{\Phi_j^2(s) \Phi_i(s)}
   \!=\!
   J_i(s)  \>.
\end{equation}
The Hartree approximation sets the third order \emph{connected} Green
function to zero:
\begin{equation}
   \frac{ \partial G_{ij}(s,s')}{\partial J_k(s'')} = 0 \>.
\end{equation}
So this means that the third order correlator is:
\begin{multline}
   \expect{\Phi_j^2(s) \, \Phi_i(s)}
   =
   \bigl [ \,
      \expect{\Phi_j(s)}^2
      +
      G_{jj}(s,s)/i \,
   \bigr ] \, \expect{\Phi_i(s)}
   \\
   +
   2 \, G_{ij}(s,s) \, \expect{\Phi_j(s)} / i \>.
\end{multline}
So the equations of motion become:
\begin{multline}
   \Bigl \{
   \bigl [ \,
      \bar{D} D
      \BM{-}
      \frac{\lambda}{2N} \,
      [ \,
         \expect{\Phi_k(s)}^2
         +
         G_{kk}(s,s)/i \,
      ]
      \BM{-}
      2 \mu \,
   \bigr ] \, \delta_{ij}
   \\
   \BM{-}
   \frac{\lambda}{N} \,
   G_{ij}(s,s)/i  \,
   \Bigr \} \, \expect{\Phi_j(s)}
   =
   0 \>,
\end{multline}
and
\begin{multline}
   \Bigl \{
   \bigl [ \,
      \bar{D} D
      \BM{-}
      \frac{\lambda}{2N} \,
      [ \,
         \expect{\Phi_l(s)}^2
         +
         G_{ll}(s,s)/i \,
      ]
      \BM{-}
      2 \mu \,
   \bigr ] \, \delta_{ij}
   \\
   \BM{-}
   \frac{\lambda}{N} \, \Phi_i(s) \Phi_j(s) \,
   \Bigr \} \, G_{jk}(s',s'')
   \\
   \BM{-}
   \frac{\lambda}{N} \,
   G_{ij}(s,s) \, G_{jk}(s,s') / i
   =
   \delta_{ik} \delta(s,s') \>.
\end{multline}
For $N=1$, these equations reduce to:
\begin{equation}\label{e:HPhi}
   \Bigl \{ \,
      \bar{D} D
      \BM{-}
      \frac{\lambda}{2} \,
      [ \,
         \Phi^2(s)
         +
         3 \, G(s,s)/i \,
      ]
      \BM{-}
      2 \mu \,
   \Bigr \} \, \Phi(s)
   =
   0 \>,
\end{equation}
and
\begin{multline}\label{e:HG}
   \Bigl \{ \,
      \bar{D} D
      \BM{-}
      \frac{3\lambda}{2} \,
      [ \,
         \Phi^2(s)
         +
         G(s,s)/i \,
      ]
      \BM{-}
      2 \mu \,
   \Bigr \} \, G(s,s')
   \\
   =
   \delta(s,s') \>.
\end{multline}
We now notice that these equations of motion are generated from an
action, given by:
\begin{multline}
   S_{H}[\Phi,\chi]
   =
   -
   \int \rd s \,
   \Bigl \{ \,
      \frac{1}{2} \, \Phi(s) \,
      \bigl [ \,
         \bar{D} D \ \BM{-} \ \chi(s) \,
      \bigr ] \, \Phi(s)
      \\
      +
      \frac{\lambda}{4} \, \Phi^4(s)
      -
      \frac{1}{3 \lambda} \,
      \bigl [ \,
         \chi^2(s)/2 \
         \JFD{-} \
         \JFD{2 \mu} \, \chi(s) \,
      \bigr ]
      \\
      +
      \frac{i}{2} \, \mathrm{Tr}
      \{ \ln [ \, \bar{D} D \ \BM{-} \ \chi(s) \, ] \}
   \Bigr \}
\end{multline}
where $\chi(s)$ is an auxiliary superfield.  The equations of motion
generated from this action is given by:
\begin{subequations}\label{e:Hartreeeqs}
\begin{gather}
   [ \, \bar{D} D \ \BM{-} \ \chi(s) - \lambda \Phi^2(s) \, ] \, \Phi(s)
   =
   0 \>,
   \label{e:Hartree.a}\\
   [ \, \bar{D} D \ \BM{-} \ \chi(s) \, ] \, G(s,s')
   =
   \delta(s,s') \>,
   \label{e:Hartree.b}\\
   \chi(s)
   =
   2 \mu \
   \BM{+} \
   (3 \lambda / 2 ) \,
   [ \, \Phi^2(s) + G(s,s)/i \, ]  \>,
   \label{e:Hartree.c}
\end{gather}
\end{subequations}
and agrees with Eqs.~\eqref{e:HPhi} and \eqref{e:HG}.  The effective
Hartree superpotential is then given by:
\begin{multline}\label{h.e:superpotential}
   V_{\text{H}}[\Phi,\chi]
   =
   \int \rd^2 \theta \,
   \Bigl \{ \,
   -
   \frac{1}{2} \, ( \bar{\theta} \cdot \theta ) \, F^2
   \ \BM{-} \
   \frac{1}{2} \, \chi \, \Phi^2
   -
   \frac{\lambda}{4} \, \Phi^4
   \\
   \BM{+}
   \frac{1}{3 \lambda} \,
   \Bigl [ \,
      \frac{\chi^2}{2} \
      \JFD{-} \
      \JFD{2 \mu} \, \chi \,
   \Bigr ]
   -
   \frac{i}{2} \,
   \Tr{ \ln G^{-1}[\Phi,\chi] } \,
   \Bigr \} \>,
\end{multline}

%
%
\subsection{Schwinger-Dyson equations and the 2-PI effective action}

We develop in this section the coupled supersymmetric Schwinger-Dyson
equations for the $O(N)$ model.  We first rewrite
Eq.~\eqref{e:largeNactionII} in an extended field scheme:
\begin{multline}\label{e:actionExt}
   \!\! S[\Phi,J]
   =
   \int \rd s \,
   \Bigl \{
   -
   \frac{1}{2} \, \int \rd s' \,
   \Phi_a(s) \,
   \Delta_{ab}^{-1}(s,s') \,
   \Phi_b(s')
   \\
   -
   \frac{1}{6} \, \gamma_{abc} \,
   \Phi_a(s) \, \Phi_b(s) \, \Phi_c(s)
   +
   J_a(s) \, \Phi_a(s) \,
   \Bigr \} \>.
\end{multline}
In the rest of this section, we have suppressed the
supercoordinates.
In this extended scheme, $a = (0,i)$, with $i = 1,2,\dotsc, N$, and
we define the extended vectors:
\begin{equation}\label{e:extendedscheme}
\begin{split}
   \Phi_a
   &=
   \bigl ( \, \chi, \Phi_1, \Phi_2, \dotsc, \Phi_N \, \bigr ) \>,
   \\
   J_a
   &=
   \bigl ( \, J_0, J_1, J_2, \dotsc, J_N \, \bigr ) \>,
\end{split}
\end{equation} where $J_0 = K + N \, \mu / \lambda$,
and
\begin{equation}
\label{dab}
   \Delta_{ab}^{-1}
   =
   \begin{pmatrix}
      D^{-1} & 0 \\
      0      & G^{-1}_{ij}
   \end{pmatrix} \>,
\end{equation}
where $D^{-1} = - N/\lambda$ and $G^{-1}_{ij} = \bar{D} \cdot D \,
\delta_{ij}$.  The introduction of a composite field $\chi$ enables us
to use a cubic supersymmetric interaction rather than the usual
quartic term, at the expense of an additional dimension for the
superfield vector $\Phi_a$.  For our case, $\gamma_{abc}$ is
\emph{fully symmetric}, and given by:
\begin{equation}
   \gamma_{0ij}
   =
   \gamma_{i0j}
   =
   \gamma_{ij0}
   =
   \delta_{ij} \>,
\end{equation}
with all other values zero.  The equation of motion for the quantum
\emph{operators} $\hat{\Phi}_a$ is
given by:
\begin{equation}\label{e:equofmotion}
   \Delta_{ab}^{-1} \, \hat{\Phi}_b
   +
   \frac{1}{2} \, \gamma_{abc} \, \hat{\Phi}_b \, \hat{\Phi}_c
   =
   J_a  \>.
\end{equation}

The supergenerating functional is given by the path integral
\eqref{e:pathintegralII}, which we write as:
\begin{equation}\label{e:pathintegralIII}
   Z[J]
   =
   e^{i W[J] }
   =
   \int \rd \Phi \, e^{i S[\Phi,J]} \>.
\end{equation}
Expectation values of the closed-time-path ordered product of $n$
field operators are given by:
\begin{equation}
   \expectTproduct{ \hat{\Phi}_a \hat{\Phi}_b \dotsc }
   =
   \frac{1}{i^n Z} \,
   \frac{\partial^n Z[J]}
        {\partial J_a \, \partial J_b \cdots } \>.
\end{equation}
The $n$-point connected supergreen functions are defined by:
\begin{equation}\label{e:Greenfundef}
   G_{ab\dotsc}[J]
   =
   \frac{\partial^n W[J]}
        {\partial J_a \, \partial J_b \cdots } \>.
\end{equation}
Here $G_{ab\dotsc}[J]$ is fully symmetric with respect to
interchange of arguments.  In particular for $n=1$:
\begin{equation}
   \Phi_a[J]
   \equiv
   G_a[J]
   =
   \frac{\partial W[J]}{\partial J_a} \>,
\end{equation}
which is the average value of the field when evaluated at $J=0$.  The
vertex function $\Gamma[\Phi]$ is defined by the Legendre
transformation:
\begin{equation}
   \Gamma[\Phi]
   =
   W[J]
   -
   J_a \, \Phi_a \>.
\end{equation}
In analogy to the supergreen functions, the $n$-point connected
supervertex functions are then defined by:
\begin{equation}\label{e:Vertexfundef}
   \Gamma_{ab\dotsc}[\Phi]
   =
   -
   \frac{\partial^n \Gamma[\Phi]}
        {\partial \Phi_a \, \partial \Phi_b \cdots } \>.
\end{equation}
In particular for $n=1$:
\begin{equation}
   J_a[\Phi]
   \equiv
   -
   \Gamma_{a}[\Phi]
   =
   -
   \frac{\partial \Gamma[\Phi]}
        {\partial \Phi_a}  \>.
\end{equation}
The two-point supergreen functional is the inverse of the two-point
supervertex functional.  Using the chain rule, we find:
\begin{equation}\label{e:identityI}
\begin{split}
   G_{ab}[J] \, \Gamma_{bc}[\Phi]
   &=
   -
   \frac{\partial^2 W[J]}
        {\partial J_a \, \partial J_b} \,
   \frac{\partial^2 \Gamma[\Phi]}
        {\partial \Phi_b \, \partial \Phi_c}
   \\
   &=
   \frac{\partial \Phi_b[J]}{\partial J_a}
   \frac{\partial J_c}{\partial \Phi_b}
   =
   \delta_{ac} \>.
\end{split}
\end{equation}
Differentiating \eqref{e:identityI} with respect to $\Phi_d$ gives:
\begin{equation*}
   \frac{\partial G_{ab}[J]}{\partial \Phi_d} \,
   \Gamma_{bc}[\Phi]
   +
   G_{ab}[J] \,
   \Gamma_{dbc}[\Phi]
   =
   0 \>.
\end{equation*}
Using \eqref{e:identityI} gives:
\begin{equation}\label{eidentityII}
   \frac{\partial G_{ae}[J]}{\partial \Phi_d}
   =
   -
   G_{ab}[J] \, \Gamma_{dbc}[\Phi] \, G_{ce}[J] \>.
\end{equation}

The Schwinger-Dyson hierarchy of coupled equations is generated by
taking the expectation value of the closed-time-path ordered product
of Eq.~\eqref{e:equofmotion}.  We find:
\begin{multline}\label{e:Gammaa}
   \Delta_{ab}^{-1} \Phi_b
   +
   \frac{1}{2} \, \gamma_{abc} \,
   \bigl ( \,
      \Phi_b \, \Phi_c
      +
      G_{bc}[J]/i \,
   \bigr )
   \\
   =
   J_a[\Phi]
   =
   -
   \Gamma_a[\Phi]
   =
   -
   \frac{\partial \Gamma[\Phi]}{\partial \Phi_a}  \>.
\end{multline}
When evaluated at $J=0$, Eq.~\eqref{e:Gammaa} is the equation of
motion for the fields $\Phi_a$.  Differentiation of
Eq.~\eqref{e:Gammaa} with respect to $\Phi_b$ gives:
\begin{equation}\label{e:Gammaab}
   \Gamma_{ab}[\Phi]
   =
   -
   \frac{\partial^2 \Gamma[\Phi]}
        {\partial \Phi_a \, \partial \Phi_b }
   =
   \bar{\Delta}_{ab}^{-1}[\Phi]
   +
   \bar{\Sigma}_{ab}[\Phi] \>,
\end{equation}
where, from \eqref{eidentityII}:
\begin{align}
   \bar{\Delta}_{ab}^{-1}[\Phi]
   &=
   \Delta_{ab}^{-1}
   +
   \gamma_{abc} \Phi_c \>,
   \label{e:barDeltabarSigma} \\
   \bar{\Sigma}_{ab}[\Phi]
   &=
   \frac{i}{2} \,
   \gamma_{aa'b'} \, G_{a'a''}[J] \, G_{b'b''}[J] \,
   \Gamma_{a''b''b}[\Phi] \>.
   \notag
\end{align}
The three-point supervertex function can now be computed by
differentiating \eqref{e:Gammaab}.  We find:
\begin{equation}\label{e:Gammaabc}
   \Gamma_{abc}[\Phi]
   =
   -
   \frac{\partial^2 \Gamma[\Phi]}
        {\partial \Phi_a \, \partial \Phi_b \, \partial \Phi_c }
   =
   \gamma_{abc}
   +
   \frac{\partial \bar{\Sigma}_{ab}[\Phi]}{\partial \Phi_c} \>.
\end{equation}
The bare vertex approximation (BVA) keeps only the first term in this
equation, in which case, we find for the self-energy:
\begin{equation}\label{e:BVAselfenergy}
   \bar{\Sigma}_{ab}^{\text{BVA}}[\Phi]
   =
   \frac{i}{2} \,
   \gamma_{aa'b'} \, G_{a'a''}[J] \, G_{b'b''}[J] \,
   \gamma_{a''b''b} \>.
\end{equation}
Using this approximation to the full self-energy, we invert
Eq.~\eqref{e:Gammaab} by multiplying by $\bar{\Delta}_{aa'}[\Phi] \,
G_{bb'}[J]$ to give the integral equation:
\begin{equation}\label{e:greeneqBVA}
   G_{ab}[J]
   =
   \bar{\Delta}_{ab}[\Phi]
   -
   \bar{\Delta}_{aa'}[\Phi] \,
   \bar{\Sigma}_{a'b'}^{\text{BVA}}[\Phi] \,
   G_{b'b}[J] \>,
\end{equation}
which is to be solved self-consistently for $G_{ab}[J]$.  The
hierarchy of coupled green function equations have now been
truncated. The BVA approximation is a \emph{conserving}
approximation in that an action can be constructed, using the
methods of Cornwall, Jackiw, and Tomboulis~\cite{r:CJT}, which
reproduces these coupled equations. This action is given by:
\begin{multline}
   S[\Phi,G]
   =
   S_{\text{class}}[\Phi]
   +
   \frac{i}{2} \,
   \mathrm{Tr} \{ \ln [ \bar{\Delta}^{-1} ] \}
   \\
   +
   \frac{i}{2} \,
   \mathrm{Tr} \{ \ln [ \bar{\Delta}^{-1}[\Phi] \, G - 1 ] \}
   +
   \Gamma_2[G]  \>,
\end{multline}
where, in the BVA,
\begin{equation}
   \Gamma_2[G]
   =
   -
   \frac{1}{12} \,
   \gamma_{abc} \, G_{aa'} \, G_{bb'} \, G_{cc'} \,
   \gamma_{a'b'c'}  \>.
\end{equation}
Varying this action with respect to $\Phi$ and $G$ independently,
leads to the BVA equations. The natural 2-PI expansion would consist
of taking higher and higher loops in $\Gamma_2$  (the lowest being
two loops).  However if one wants to further keep only terms in
$\Gamma_2$ to a particular order in $1/N$ then one needs to realize
that the $\chi$ and $\phi$ pieces of $G$ have different $N$
dependence as seen in Eqs.~\eqref{e:extendedscheme} and~\eqref{dab}.
This is discussed in detail in Ref.~\cite{r:AABBS}

%
%
\section{Effective potential}

In this section, we derive effective potentials for the large-$N$ and
Hartree approximations in the vacuum at $T=0$ and for finite
temperature.

We consider here the spatially homogeneous case where the average
superfields depend only on time, and require the average Fermi field
to vanish.  Thus we write:
\begin{align}
   \Phi_i(t,\theta)
   &=
   \phi_i(t)
   +
   \frac{1}{2} \, \bar{\theta} \cdot \theta \, F_i(t) \>,
   \label{e:Phiexpanded} \\
   \chi(t,\theta)
   &=
   2 \, \rho(t) \
   \BM{-} \
   \bar{\theta} \cdot \theta \, R(t) \>.
   \label{e:chiexpanded}
\end{align}

%
%
\subsection{Supergreen function}

The two-point supergreen function $G(s,s')$ is of the form:
\begin{multline}\label{e:Gexpanded}
   G_{ij}(s,s')
   =
   g_{0\,ij}(x,x')
   +
   \frac{1}{2} \,
   ( \,
      \bar{\theta} \cdot \theta
      +
      \bar{\theta}' \cdot \theta' \,
   ) \, g_{1\,ij}(x,x')
   \\
   \BM{-}
   \bar{\theta} \cdot g_{2\,ij}(x,x') \cdot \theta'
   +
   \frac{1}{4} \,
   ( \bar{\theta} \cdot \theta ) \,
   ( \bar{\theta}' \cdot \theta' ) \,
   g_{3\,ij}(x,x') \>.
\end{multline}
The generalized Ward-Takahashi identity states that:
\begin{equation}\label{e:WTidentity}
   ( \, Q + Q' \, ) \, G_{ij}(s,s') = 0  \>.
\end{equation}
Here $Q$ and $Q'$ are the supercharge operators, given by:
\begin{equation}
   Q
   =
   -
   \bar{\partial}
   -
   i \, \pslash \cdot \theta  \>,
   \qquad
   Q'
   =
   -
   \partial'
   -
   i \, \pslash' \cdot \theta' \>.
\end{equation}
Eq.~\eqref{e:WTidentity} requires that:
\begin{subequations}\label{e:WTaall}
\begin{align}
   g_{2\,ij}(x',x)
   &=
   -
   g_{1\,ij}(x,x')
   -
   i \, \pslash g_{0\,ij}(x,x') \>,
   \label{e:WTai}\\
   g_{2\,ij}(x,x')
   &=
   -
   g_{1\,ij}(x,x')
   -
   i \, \pslash' g_{0\,ij}(x,x') \>,
   \label{e:WTaii}\\
   g_{3\,ij}(x,x')
   &=
   -
   i \, \pslash g_{1\,ij}(x,x')
   +
   i \, \pslash' g_{2\,ij}(x',x) \>,
   \label{e:WTaiii}\\
   g_{3\,ij}(x,x')
   &=
   -
   i \, \pslash' g_{1\,ij}(x,x')
   +
   i \, \pslash g_{2\,ij}(x,x') \>,
   \label{e:WTaiv}
\end{align}
\end{subequations}
from which we obtain:
\begin{equation}\label{e:WTI}
   g_{3\,ij}(x,x')
   =
   \partial^{\mu} \partial'_{\mu} \, g_{0\,ij}(x,x') \>,
\end{equation}
and
\begin{multline}\label{e:WTII}
   \frac{1}{2} \,
   \bigl \{ \,
      \bar{\theta} \cdot g_{2\,ij}(x,x') \cdot \theta'
      +
      \bar{\theta}' \cdot g_{2\,ij}(x',x) \cdot \theta \,
   \bigr \}
   \\
   =
   -
   ( \bar{\theta} \cdot \theta' ) \, g_{1\,ij}(x,x')
   +
   \bar{\theta} \cdot
   \frac{i}{2} \,
   \bigl [ \,
      \pslash - \pslash' \,
   \bigr ] \cdot \theta' \, g_{0\,ij}(x,x') \>.
\end{multline}
So using \eqref{e:WTI} and \eqref{e:WTII}, if $G(s,s')$ satisfies the
Ward-Takahashi identity, it must be of the general form:
\begin{align}\label{e:Gexpandedresult}
   G(s,s')
   = &
   \Bigl \{
      1
      +
      \frac{i}{2}
      \bar{\theta} \cdot
      \bigl [
         \Slash{\partial} - \Slash{\partial}'
      \bigr ] \cdot
      \theta'
   \\ \notag & \quad
      +
      \frac{1}{4}
      ( \bar{\theta} \cdot \theta ) \cdot
      ( \bar{\theta}' \cdot \theta' )
      \partial^{\mu} \partial'_{\mu}
   \Bigr \} \, g_{0\,ij}(x,x')
   \\ \notag &
   +
   \frac{1}{2} \,
   \Bigl \{ \,
      \bar{\theta} \cdot \theta
       -
       2 \, \bar{\theta} \cdot \theta'
       +
       \bar{\theta}' \cdot \theta' \,
   \Bigr \} \, g_{1\,ij}(x,x')
\end{align}
or
\begin{align}
   G(s,s')
   = &
   \exp
   \Bigl \{ \,
      \frac{i}{2} \,
      \bar{\theta} \cdot
      \bigl [ \,
         \pslash - \pslash' \,
      \bigr ] \cdot
      \theta' \,
   \Bigr \} \,
   g_{0\,ij}(x,x')
   \notag \\ &
   +
   \frac{1}{2} \,
   \delta^2(\theta - \theta') \,
   g_{1\,ij}(x,x') \>.
\end{align}
The supergreen function satisfies an equation of the form:
\begin{equation}\label{e:greeneq}
   \bigl [ \,
      \bar{D} \cdot D \
      \BM{-} \
      \chi(t) \,
   \bigr ] \, G_{ij}(s,s')
   =
   \delta_{ij} \delta(s,s') \>,
\end{equation}
from which we find the component equations:
\begin{equation}\label{e:greencomeqs}
\begin{split}
   \bigl [ \,
      \Box + m^2(t) \,
   \bigr ] \, g_{0\,ij}(x,x')
   &=
   \delta_{ij} \delta(x,x') \>,
   \\
   \bigl [ \,
      i \, \pslash
      -
      \rho(t) \,
   \bigr ] \, g_{2\,ij}(x,x')
   &=
   \delta_{ij} \delta(x,x') \>,
\end{split}
\end{equation}
with $m^2(t) = \rho^2(t) + R(t)$.  $g_{1\,ij}(x,x')$ and
$g_{3\,ij}(x,x')$ can be found from Eqs.~\eqref{e:WTaall}.  We will
use these results below.

%
%
\subsection{Large-$N$ approximation}
\label{s:eomlargeN}

In the large-$N$ approximation, the gap equation \eqref{e:folNiii}
becomes:
\begin{subequations}\label{e:gapcomponents}
\begin{align}
   \rho(t)
   &=
   \mu
   +
   ( \lambda / 4 N ) \,
   \bigl [ \,
      \phi_i^2(t)
      +
      g_{0\,ii}(t,t) / i \,
   \bigr ] \>,
   \label{e:gaprho} \\
   R(t)
   &=
   ( \lambda / 2 N ) \,
   \bigl \{ \,
      \rho(t) \,
      \bigl [ \,
         \phi_i^2(t)
         +
         g_{0\,ii}(t,t) / i \,
      \bigr ]
      \notag \\
      &\qquad\qquad\qquad\qquad
      +
      \Tr{ g_{2\,ii}(t,t) } / i \,
   \bigr \} \>.
   \label{e:gapR}
\end{align}
\end{subequations}
Here, we have used $F_i(t) = \rho(t) \phi_i(t)$.

In the vacuum where $\Phi_i(\theta)$ and $\chi(\theta)$ depend only on
$\theta$, the supergreen function can easily be computed in terms of
$\rho$ and $R$.  Performing a Wick rotation to Euclidean coordinates,
we set:
\begin{equation}
   G_{ij}(s,s')/i
   =
   \int \frac{\rd^3 k}{(2\pi)^3} \,
   \tilde{G}_{ij}(k;\theta,\theta') \,
   e^{-i k \cdot ( x - x' )} \>.
\end{equation}
Using Eq.~\eqref{e:greeneq}, we find $\tilde{G}_{ij}(k;\theta,\theta')
= \delta_{ij} \, \tilde{G}(k;\theta,\theta')$, with :
\begin{align}\label{lN.e:fgreen}
   & \tilde{G}(k;\theta,\theta')
   =
   \frac{
      \bar{\theta} \cdot
      ( \, i \kslash - \rho \, )
      \cdot \theta'
        }{ k^2 + \rho^2 }
   \\ \notag &
   + \frac{
      1
      +
      \frac{1}{2}
      ( \bar{\theta} \cdot \theta
           +
           \bar{\theta}' \cdot \theta'
      ) \rho
      -
      \frac{1}{4}
      ( \bar{\theta}  \cdot \theta  )
      ( \bar{\theta}' \cdot \theta' ) ( k^2 + \rho^2 )
        }{ k^2 + \rho^2 + R }
    \>.
\end{align}
The diagonal elements are given by:
\begin{equation}\label{lNre.e:Gdiag}
   \tilde{G}_{ii}(k,\theta,\theta)
   =
   \frac{ 1 + \bar{\theta} \cdot \theta \, \rho }
        { k^2 + \rho^2 + R }
   -
   \frac{ \bar{\theta} \cdot \theta \, \rho }{ k^2 + \rho^2 } \>,
\end{equation}
independent of $i$.  We identify the Boson mass with $m = \sqrt{
\rho^2 + R}$ and the Fermion mass with $\rho$.  The gap equations
\eqref{e:gapcomponents} become:
\begin{subequations}\label{e:gapsvacuum}
\begin{align}
   \rho
   &=
   \mu
   +
   \frac{\lambda}{4}
   \Bigl [
      \phi^2
      +
      \int^{\Lambda} \frac{\rd^3 k}{(2\pi)^3}
      \frac{1}{ k^2 + \rho^2 + R }
   \Bigr ] \>,
   \label{e:gapsva} \\
   R
   &=
   \frac{\lambda}{2} \, \rho \,
   \Bigl \{
      \phi^2
      +
      \int \frac{\rd^3 k}{(2\pi)^3}
      \Bigl [
         \frac{1}{ k^2 + \rho^2 + R }
         -
         \frac{1}{ k^2 + \rho^2 }
      \Bigr ]
   \Bigr \}  \>.
   \label{e:gapsvb}
\end{align}
\end{subequations}
Here we have introduced a three-dimensional cutoff $\Lambda$ to make
Eq.~\eqref{e:gapsva} finite.  Due to the magic of supersymmetry,
Eq.~\eqref{e:gapsvb} is finite.

%
%

For $d=3$ we renormalize Eq.~\eqref{e:gapsva} by subtracting it
about the point $k^2 = 0$ with a renormalized constant $\mu_R$
defined by:
\begin{equation}\label{lNd3.e:muR}
   \mu_R
   =
   \mu
   +
   \frac{\lambda}{4}
   \int^{\Lambda} \frac{\rd^3 k}{(2\pi)^3} \,
      \frac{1}{ k^2 } \>.
\end{equation}
This gives the renormalized gap equation:
\begin{align}
   \rho
   &=
   \mu_R
   +
   \frac{\lambda}{4} \,
   \Bigl ( \,
      \phi^2
      -
      \frac{1}{4\pi} \, | m | \,
   \Bigr ) \>,
   \label{lN3d.e:gapsvarenorm}
\end{align}
where $m^2 = \rho^2 + R$.  Eq.~\eqref{e:gapsvb} is finite, and
yields:
\begin{equation}\label{lN3d.e:gapsvbrenorm}
   R
   =
   \frac{\lambda}{2} \, \rho \,
   \Bigl [ \,
      \phi^2
      -
      \frac{1}{4\pi} \,
      ( \, | m | - | \rho | \, ) \,
   \Bigr ] \>.
\end{equation}
Multiplying Eq.~\eqref{lN3d.e:gapsvarenorm} by $2 \rho$, and
subtracting it from Eq.~\eqref{lN3d.e:gapsvbrenorm} gives:
\begin{equation}\label{lN3d.e:rhotoR}
   R
   =
   2 \, \rho \, ( \, \rho - \mu_R \, )
   -
   \frac{\lambda}{8\pi} \, \rho \, | \rho | \>,
\end{equation}
which relates $R$ to $\rho$.

From Eq.~\eqref{lN.e:superpotential}, and using
\eqref{e:Phiexpanded} and \eqref{e:chiexpanded}, and the
renormalization prescription \eqref{lNd3.e:muR}, the large-$N$
effective potential for $N=1$ is given by:
\begin{equation}\label{lNep.e:VNi}
   V_N(\phi,F,\rho,R)
   =
   V_{\text{c}}(\phi,F,\rho,R)
   +
   V_{\text{q}}(\rho,R) \>,
\end{equation}
where the classical part is given by:
\begin{multline}\label{lNep.e:Vcli}
   V_{\text{c}}(\phi,F,\rho,R)
   =
   \\
   \rho \, \phi \, F
   -
   \frac{1}{2} \, F^2
   +
   \frac{1}{2} \, R \, \phi^2
   +
   \frac{2}{\lambda} \, R \, ( \, \mu_R - \rho \, )
\end{multline}
and the quantum part by:
\begin{equation}\label{lNep.e:Vpi}
   V_{\text{q}}(\rho,R)
   =
   \int \rd^2 \theta \, W_{\text{q}}(\rho,R,\theta) \>,
\end{equation}
with
\begin{multline}
   W_{\text{q}}(\rho,R,\theta)
   =
   \\
   \frac{1}{2}
   \int \frac{\rd^3 k}{(2\pi)^3} \,
   \Bigl \{ \,
      \ln [ \, G^{-1}(k;\theta,\theta) \, ]
      -
      \frac{\chi(\theta)}{ k^2 } \,
   \Bigr \} \>.
\end{multline}
At the minimum of the potential, $F = \rho \, \phi$.  Evaluating
\eqref{lNep.e:Vcli} at this value of $F$ yields:
\begin{equation}\label{lNd3.e:Vclii}
   V_{\text{c}}(\phi,\rho,R)
   =
   \frac{1}{2} \, m^2 \, \phi^2
   +
   \frac{2}{\lambda} \, R \, ( \, \mu_R - \rho \, )
   \>,
\end{equation}
where we have again set $m^2 = \rho^2 + R$.  For
$W_{\text{q}}(\rho,R)$, it is easier to first evaluate:
\begin{equation}\label{lNep.e:dVp}
\begin{split}
   \frac{\delta W_{\text{q}}(\rho,R,\theta)}{\delta \chi(\theta)}
   &=
   \frac{1}{2}
   \int \frac{\rd^3 k}{(2\pi)^3} \,
   \Bigl [ \,
      \tilde{G}(k;\theta,\theta)
      -
      \frac{1}{ k^2 } \,
   \Bigr ]
   \\
   &=
   W_0(\rho,R)
   +
   \frac{1}{2} \, \bar{\theta} \cdot \theta \,
   W_1(\rho,R) \>.
\end{split}
\end{equation}
Using \eqref{lNre.e:Gdiag}, we find:
\begin{equation}\label{Lnep.e:W0}
\begin{split}
   W_0(\rho,R)
   &=
   \frac{1}{2}
   \int \frac{\rd^3 k}{(2\pi)^3} \,
   \Bigl ( \,
      \frac{1}{ k^2 + \rho^2 + R }
      -
      \frac{1}{ k^2 } \,
   \Bigr )
   \\
   &=
   -
   \frac{1}{8\pi} \, | m | \>,
\end{split}
\end{equation}
and
\begin{equation}\label{Lnep.e:W1}
\begin{split}
   W_1(\rho,R)
   &=
   \int \frac{\rd^3 k}{(2\pi)^3} \,
   \Bigl ( \,
      \frac{\rho}{ k^2 + \rho^2 + R }
      -
      \frac{\rho}{ k^2 + \rho^2 } \,
   \Bigr )
   \\
   &=
   -
   \frac{\rho}{4\pi} \, ( \, | m | - | \rho | \, ) \>.
\end{split}
\end{equation}
Now since $\delta \chi( \rho, R ) = 2 \delta \rho + \bar{\theta}
\cdot \theta \, \delta R$, we have:
\begin{multline}
   \delta W_{\text{q}}(\rho,R,\theta)
   =
   \\
   \bigl [ \,
      W_0(\rho,R)
      +
      \frac{1}{2} \, \bar{\theta} \cdot \theta \,
      W_1(\rho,R) \,
   \bigr [ \,
   \bigl ( \,
      2 \delta \rho
      +
      \bar{\theta} \cdot \theta \, \delta R \,
   \bigr )
   \\
   =
   2 \, W_0(\rho,R) \, \delta \rho
   \\
   +
   \bar{\theta} \cdot \theta \,
   \bigl [ \,
      W_0(\rho,R) \, \delta R
      +
      W_1(\rho,R) \, \delta \rho \,
   \bigr ] \>.
\end{multline}
For the effective potential, we only need the last term.  So we now
want to find a common function $V_{\text{q}}(\rho,R)$ such that:
\begin{equation*}
   \frac{\partial V_{\text{q}}(\rho,R)}{\partial R}
   =
   W_0(\rho,R) \>,
   \quad
   \frac{\partial V_{\text{q}}(\rho,R)}{\partial \rho}
   =
   W_1(\rho,R) \>.
\end{equation*}
Such a function is given by:
\begin{equation}\label{lNd3.e:Vqii}
   V_{\text{q}}(\rho,R)
   =
   -
   \frac{1}{12 \pi} \,
   \bigl ( \,
      | m |^3
      -
      | \rho |^3 \,
   \bigr ) \>.
\end{equation}
So from \eqref{lNd3.e:Vclii} and \eqref{lNd3.e:Vqii} the effective
potential is given by:
\begin{multline}\label{lNd3.e:VNi}
   V_N(\phi,\rho,R)
   =
   \frac{1}{2} \, m^2 \, \phi^2
   +
   \frac{2}{\lambda} \, R \, ( \, \mu_R - \rho \, )
   \\
   -
   \frac{1}{12 \pi} \,
   \bigl ( \,
      | m |^3
      -
      | \rho |^3 \,
   \bigr ) \>.
\end{multline}
The minimum of the potential is at the point $(\phi,\rho,R)$ defined
by the equations:
\begin{equation*}
   \frac{\partial V_N(\phi,\rho,R)}{\partial \phi}
   =
   \frac{\partial V_N(\phi,\rho,R)}{\partial \rho}
   =
   \frac{\partial V_N(\phi,\rho,R)}{\partial R}
   =
   0 \>.
\end{equation*}
The first partial derivative gives the requirement:
\begin{equation}\label{lNd3.e:Vphi}
   m^2 \, \phi = 0 \>,
\end{equation}
so the minimum of the potential is at either $m^2 = 0$ or $\phi =
0$. The last two partial derivatives gives the two gap equations:
\begin{subequations}\label{lNd3.e:gapeqs}
\begin{align}
   \rho
   &=
   \mu_R
   +
   \frac{\lambda}{4} \,
   \Bigl ( \,
      \phi^2
      -
      \frac{1}{4\pi} \, | m | \,
   \Bigr ) \>,
   \label{lNd3.e:gaprho}  \\
   R
   &=
   \frac{\lambda}{2} \, \rho \,
   \Bigl [ \,
      \phi^2
      -
      \frac{1}{4\pi} \,
      \bigl ( \, | m | - | \rho | \, \bigr ) \,
   \Bigr ] \>.
   \label{lNd3.e:gapR}
\end{align}
\end{subequations}
Using \eqref{lNd3.e:gaprho} to eliminate the $2 ( \mu_R - \rho ) /
\lambda$ term in Eq.~\eqref{lNd3.e:VNi}, we find that \emph{at the
minimum of the potential,}
\begin{multline}\label{VN}
   V_N(\phi,\rho,m)
   =
   \frac{1}{2} \, \rho^2 \, \phi^2
   \\
   +
   \frac{1}{24\pi} \,
   \bigl ( \, | m | - | \rho | \, \bigr )^2 \,
   ( \, | m | + 2 \, | \rho | \, ) \>,
\end{multline}
in agreement with MZJ \cite{ref:MMZJ}[Eq.~(2.19)].  For any value of
$\phi$, the minimum of the potential is when $| m | = | \rho |$,
that is when $R = 0$, in which case either $\phi = 0$ or $ \rho = m
= 0$. In both cases, $V_N(\phi,\rho,m) = 0$.

\begin{figure*}[t]
   \centering
   \includegraphics[width=0.8\textwidth]{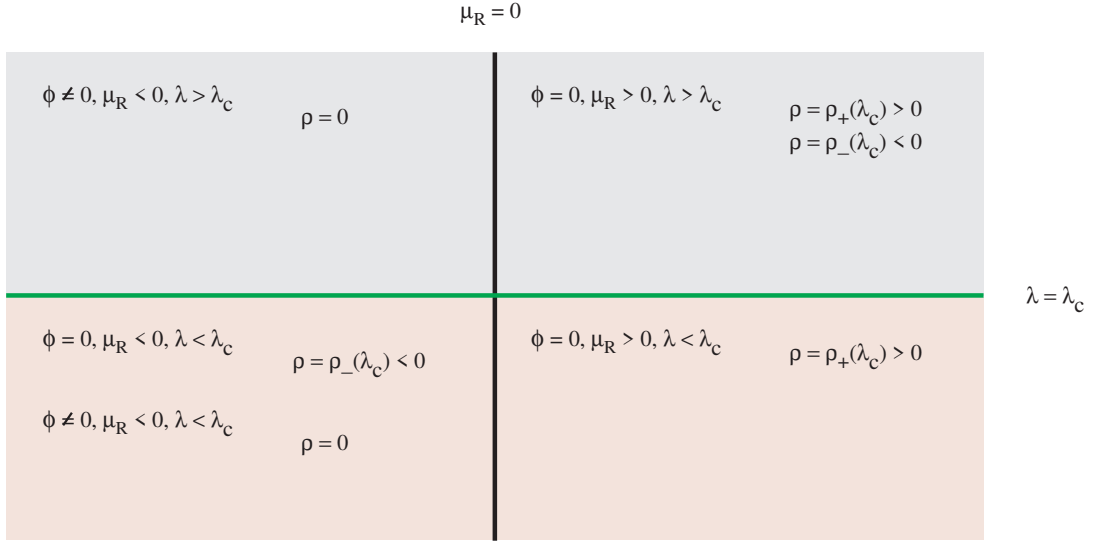}
   \caption{(Color online) Phase structure of the supersymmetric
   $O(N)$ model at zero temperature in the large-$N$ approximation.
   We plot $\mu_R$ on the $x$-axis and $\lambda >0$ on the $y$-axis,
   and have defined $\rho_\pm(\lambda_c) = \mu_R/\bigl ( 1 \pm
   \lambda/\lambda_c \bigr )$.}
   \label{f:lNphase}
\end{figure*}

A given renormalized theory is specified by the parameters $\mu_R$ and
$\lambda$.  We therefore have the following possibilities:
\begin{enumerate}

\item When $\phi = 0$ (the unbroken symmetry case), at the minimum of
  the potential, $\rho$ must satisfy Eq.~\eqref{lNd3.e:gaprho}:
\begin{equation}\label{lN3d.e:gaprhophizero}
   \rho = \mu_R - | \, \rho \, | \, ( \, \lambda / \lambda_c \, )\>,
\end{equation}
  where we have set $\lambda_c = 16 \pi$.  If $\rho > 0$ then we have
  that
\begin{equation}\label{e:rho1}
   \rho = \mu_R / ( \, 1 + \lambda / \lambda_c \, ) \>,
\end{equation}
  which is satisfied for $\mu_R>0$ and for all $\lambda >0$, with mass
  $\rho$ given by \eqref{e:rho1}. If $\rho < 0$ then from
  \eqref{lN3d.e:gaprhophizero}, we have that
\begin{equation}\label{e:rho2}
   \rho = \mu_R / ( \, 1 - \lambda / \lambda_c \, ) \>,
\end{equation}
  which can be satisfied in two ways: either (a) $\mu_R>0$ and
  $\lambda > \lambda_c$, in which case the vacuum is degenerate with
  $\rho$ masses given by Eqs.~\eqref{e:rho1} and~\eqref{e:rho2}, or
  (b) $\mu_R < 0$ and $\lambda_c > \lambda > 0$, with $\rho$ mass
  given by \eqref{e:rho2}.

\item When $\phi \neq 0$ (the broken symmetry case), then $\rho = m =
  0$, which leads to the constraint:
\begin{equation}
   \mu_R + \frac{\lambda}{4}\, \phi^2 = 0 \>.
\end{equation}
  Since $\lambda > 0$, this means that broken symmetry can occur only
  when $\mu_R < 0$.  The broken symmetry vacuum will be degenerate
  with the symmetric vacuum when $\lambda_c > \lambda > 0$.

\end{enumerate}
In all cases, the effective potential $V_N = 0$ at the minimum.  We
summarize these large-$N$ results in Fig.~\ref{f:lNphase}.

The effective potential at finite temperature is worked out in
Appendix~\ref{tg.s:tempgreen}.  From Eq.~\eqref{a.tg3d.e:Veff}, we
have:
\begin{multline}\label{tg3d.e:Veff}
   V_N(\phi,\rho,R;\beta)
   =
   \frac{1}{2} \, m^2 \, \phi^2
   +
   \frac{2}{\lambda} \, R \, ( \, \mu_R - \rho \, )
   \\
   -
   \frac{1}{12\pi}
   \bigl ( \, | m |^3 - | \rho |^3 \, \bigr )
   \\
   +
   \frac{1}{\beta}
   \int_{0}^{+\infty} \frac{k \, \rd k}{2 \pi} \,
   \ln
   \biggl [ \,
      \frac{1 - \exp ( - \beta \omega_k )}
           {1 + \exp ( - \beta \omega'_k )}
   \biggr ] \>.
\end{multline}
At the minimum of the potential, $\rho$ and $R$ satisfy the gap
equations \eqref{tgd3.e:trho} and \eqref{tgd3.e:tR} at finite
temperature:
\begin{subequations}\label{lN3d.e:gapT}
\begin{align}
   \rho
   &=
   \mu_R
   +
   \frac{\lambda}{4}
   \Bigl \{
      \phi^2
      -
      \frac{ | m | }{4\pi}
      \label{lN3d.e:rhogapT}
      -
      \frac{1}{2\pi\beta}
      \ln \bigl [ 1 - \exp(-\beta | m | ) \bigr ]
   \Bigl \}
   \\
   R
   &=
   \frac{\lambda}{2} \, \rho \,
   \Bigl \{ \,
      \phi^2
      -
      \frac{1}{2\pi\beta} \,
      \Bigl \{ \,
         \ln \bigl [ \, 2 \sinh(\beta | m |/2) \bigr ]
         \notag \\ & \qquad\qquad
         -
         \ln \bigl [ \, 2 \cosh(\beta |\rho|/2) \bigr ] \,
      \Bigr \} \,
   \Bigr \} \>.
   \label{lN.3d.RgapT}
\end{align}
\end{subequations}
Noting that $R = m^2 - \rho^2$, these two equations can be combined
to give $m^2$ as a function of $\rho$\ :
\begin{align}
   m^2
   &=
   3 \, \rho^2
   -
   2 \, \rho \, \mu_R
\label{lN.3d:rhotom}
   \\ \notag & \qquad
   +
   \frac{\lambda}{8\pi} \, | \rho | \, \rho
   +
   \frac{\lambda}{4\pi\beta} \,
   \ln \bigl [ \, 1 + \exp( - \beta | \rho | ) \, \bigr ] \>,
\end{align}
so that \emph{at the minimum}, the potential can be written as:
\begin{align}\label{tg3d.e:Veffi}
   V_N& (\phi,\rho,m;\beta)
   \!=\!
   \frac{1}{2} \rho^2 \phi^2
   \!+\!
   \frac{1}{24\pi}
   \bigl ( \, | m |  \!-\! | \rho | \, \bigr )^2
   ( | m | \!+\! 2 | \rho | )
   \notag \\ &
   +
   \frac{m^2 - \rho^2}{4\pi\beta} \,
   \ln \bigl [ \, 1 - \exp(-\beta | m | ) \, \bigr ]
   \notag \\ &
   +
   \frac{1}{\beta}
   \int_{0}^{+\infty} \frac{k \, \rd k}{2 \pi} \,
   \ln
   \biggl [ \,
      \frac{1 - \exp ( - \beta \omega_k )}
           {1 + \exp ( - \beta \omega'_k )}
   \biggr ] \>.
\end{align}

%
%
\subsection{Hartree approximation}
\label{s:eomHartree}

For the Hartree approximation, the gap equation~\eqref{e:Hartree.c}
becomes:
\begin{subequations}\label{h.e:gapcomponents}
\begin{align}
   \rho(t)
   &=
   \mu
   +
   3 \lambda / 4 \,
   \bigl [ \,
      \phi^2(t)
      +
      g_{0}(t,t) / i \,
   \bigr ] \>,
   \label{h.e:gaprho} \\
   R(t)
   &=
   3 \lambda / 2  \,
   \Bigl \{ \,
      \rho(t) \,
      \bigl [ \,
         \phi^2(t)
         +
         g_{0}(t,t) / i \,
      \bigr ] - \frac{\lambda}{2}\phi^4
      \notag \\
      &\qquad\qquad\qquad
      +
      \Tr{ g_{2}(t,t) } / i \,
   \Bigr \} \>.
   \label{h.e:gapR}
\end{align}
\end{subequations}
Here, we have set $N=1$.
The Green function in the vacuum is the same as in the large-$N$
approximation, and is given by Eq.~\eqref{lN.e:fgreen}.  So the gap
equations in the vacuum are given by:
\begin{subequations}\label{h.e:gapsvacuum}
\begin{align}
   \rho
   &=
   \mu
   +
   \frac{3\lambda}{4} \,
   \Bigl [ \,
      \phi^2
      +
      \int^{\Lambda} \frac{\rd^3 k}{(2\pi)^3} \,
      \frac{1}{ k^2 + \rho^2 + R } \,
   \Bigr ] \>,
\label{h.e:gapsva}
   \\
   R & =
   \frac{3\lambda}{2} \rho
\label{h.e:gapsvb}
   \Bigl [
      \phi^2 - \frac{\lambda}{2} \, \phi^4
   \\ \notag & \qquad
      +
      \int \frac{\rd^3 k}{(2\pi)^3} \,
      \Bigl (
         \frac{1}{ k^2 + \rho^2 + R }
         -
         \frac{1}{ k^2 + \rho^2 }
      \Bigr )
   \Bigr ]  \>.
\end{align}
\end{subequations}

%
%

We renormalize Eq.~\eqref{h.e:gapsva} by subtracting it
about the point $k^2 = 0$ with a renormalized constant $\mu_R$
defined by:
\begin{equation}\label{hd3.e:muR}
   \mu_R
   =
   \mu
   +
   \frac{3\lambda}{4}
   \int^{\Lambda} \frac{\rd^3 k}{(2\pi)^3} \,
      \frac{1}{ k^2 } \>.
\end{equation}
This gives the renormalized gap equation:
\begin{align}
   \rho
   &=
   \mu_R
   +
   \frac{3\lambda}{4} \,
   \Bigl ( \,
      \phi^2
      -
      \frac{1}{4\pi} \, | m | \,
   \Bigr ) \>,
   \label{hd3.e:gapsvarenorm}
\end{align}
where $m^2 = \rho^2 + R$.  Eq.~\eqref{h.e:gapsvb} is finite, and
yields:
\begin{equation}\label{hd3.e:gapsvbrenorm}
   R
   =
   \frac{3\lambda}{2} \, \rho \,
   \Bigl [ \,
      \phi^2
      -
      \frac{1}{4\pi} \,
      ( \, | m | - | \rho | \, ) \,
   \Bigr ]
   - \frac{3\lambda^2}{4} \, \phi^4
   \>.
\end{equation}

\begin{figure*}[t]
   \centering
   \includegraphics[width=0.8\textwidth]{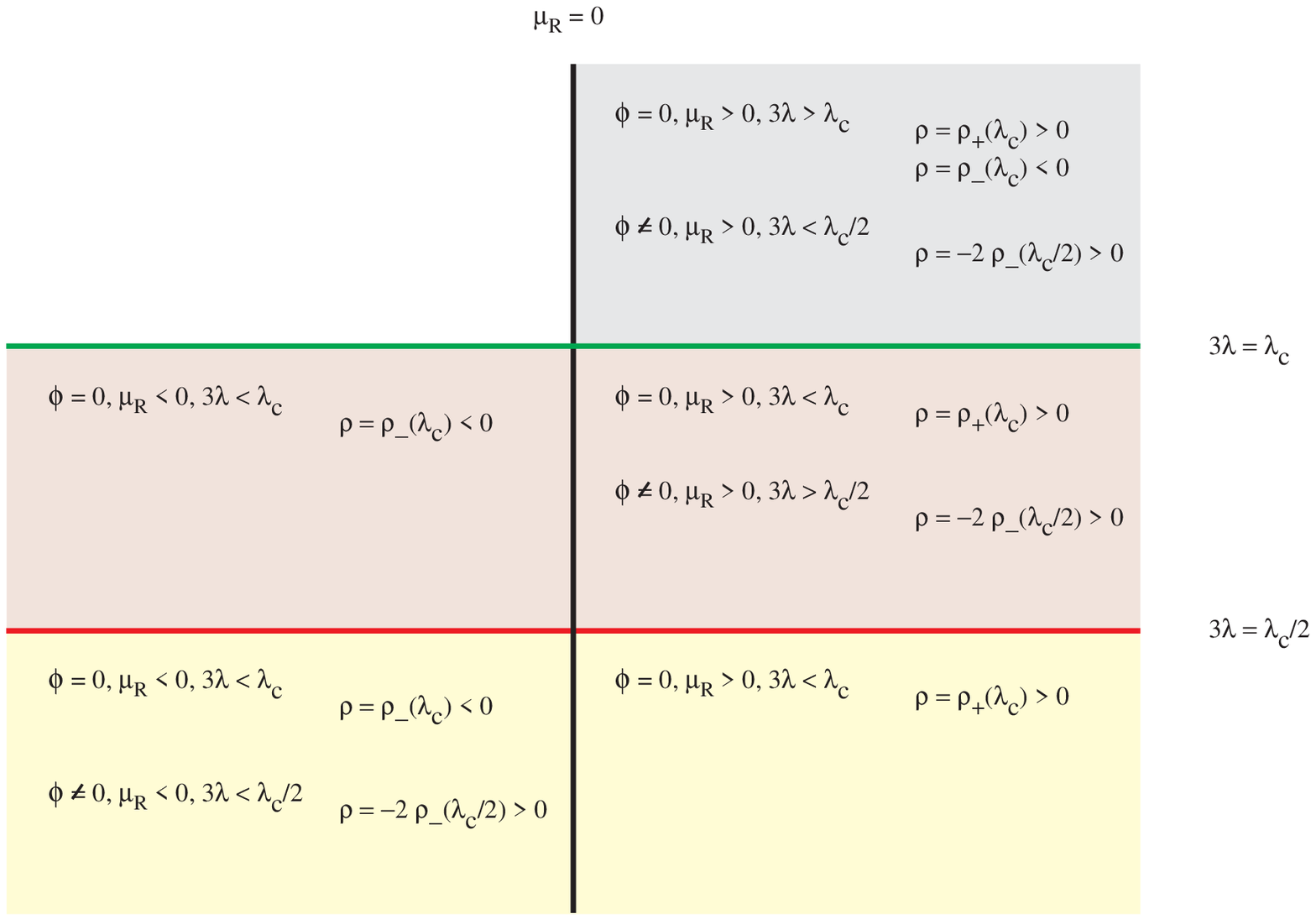}
   \caption{(Color online) Phase structure of the supersymmetric
   $O(N)$ model at zero temperature in the Hartree approximation.  We
   plot $\mu_R$ on the $x$-axis and $\lambda >0$ on the $y$-axis, and
   have defined here, we use the notation $\rho_\pm(\lambda_c) =
   \mu_R/\bigl ( 1 \pm 3\lambda/\lambda_c \bigr )$.}
   \label{f:phase}
\end{figure*}

From Eq.~\eqref{h.e:superpotential}, and using \eqref{e:Phiexpanded}
and \eqref{e:chiexpanded}, and the renormalization prescription
\eqref{hd3.e:muR}, the Hartree effective potential for $N=1$ is
given by:
\begin{equation}\label{hd3.e:VNi}
   V_H(\phi,F,\rho,R)
   =
   V_{\text{c}}(\phi,F,\rho,R)
   +
   V_{\text{q}}(\rho,R) \>,
\end{equation}
The classical part is now given by
\begin{multline}\label{H.e:Vcli}
   V_{\text{c}}(\phi,F,\rho,R)
   =
   \rho \, \phi \, F
   -
   \frac{1}{2} \, F^2
   +
   \frac{1}{2} \, R \, \phi^2
   \\
   -
   \frac{\lambda}{2} \, F \,\phi^3
   +
   \frac{2}{3\lambda} \, R \, ( \, \mu_R - \rho \, ) \>.
\end{multline}
The quantum part is the same as in the large-$N$ case and is given by
Eq.~\eqref{lNd3.e:Vqii}.  So the effective potential in the vacuum for
the Hartree approximation is given by:
\begin{multline}\label{Hd3.e:effVii}
   V_H(\phi,F,\rho,R)
   =
   \rho \, \phi \, F
   -
   \frac{1}{2} \, F^2
   +
   \frac{1}{2} \, R \, \phi^2
   -
   \frac{\lambda}{2} \, F \,\phi^3
   \\
   +
   \frac{2}{3\lambda} \, R \, ( \, \mu_R - \rho \, )
   -
   \frac{1}{12 \pi} \,
   \bigl ( \,
      | m |^3
      -
      | \rho |^3 \,
   \bigr ) \>.
\end{multline}
The minimum of the potential is when
\begin{equation}\label{Hd2.e:Fmin}
   F
   =
   \rho \, \phi - \frac{\lambda}{2} \, \phi^3 \>,
\end{equation}
at which point, the effective potential is given by:
\begin{multline}\label{Hd3.e:effViii}
   V_H(\phi,\rho,R)
   =
   \frac{1}{2} \, m^2 \, \phi^2
   -
   \frac{\lambda}{2} \, \rho \, \phi^4
   +
   \frac{\lambda^2}{8} \, \phi^6
   \\
   +
   \frac{2}{3\lambda} \, R \, ( \, \mu_R - \rho \, )
   -
   \frac{1}{12 \pi} \,
   \bigl ( \,
      | m |^3
      -
      | \rho |^3 \,
   \bigr ) \>.
\end{multline}
Minimizing with respect to $\rho$ and $R$ again give the gap
equations, \eqref{hd3.e:gapsvarenorm} and
\eqref{hd3.e:gapsvbrenorm}. At the minimum, the effective potential
can be written as
\begin{multline}\label{hartree_T0}
   V_H(\phi,\rho)
   =
   \frac{1}{2} \, \rho^2 \, \phi^2
   -
   \frac{\lambda}{2} \, \rho \, \phi^4
   +
   \frac{\lambda^2}{8} \, \phi^6
   \\
   +
   \frac{1}{24\pi}
   ( \, |m| - |\rho| \, )^2 \,
   ( \, |m| + 2 \, |\rho| \, )
   \>,
\end{multline}
where we note the terms proportional to $\phi^4$ and $\phi^6$
present in the Hartree potential in contrast with the leading-order
large-$N$ result (see Eq.~\eqref{VN}).

The minimum of~\eqref{hartree_T0} with respect to $\phi$ occurs when
\begin{equation}
   \frac{\partial V_H(\phi,\rho)}{\partial \phi}
   =
   \Bigl \{ \,
      \rho^2 - 2 \lambda \rho \, \phi^2
      + \frac{3}{4} \lambda^2 \phi^4 \,
   \Bigr \} \, \phi
   =
   0 \>,
\end{equation}
which gives the solutions $\phi = 0$ and $\phi = \phi_{\pm}$, where:
\begin{equation}
   \phi_+^2
   =
   \frac{2|\rho|}{\lambda}
   \>, \qquad
   \phi_-^2
   =
   \frac{2|\rho|}{3\lambda}
\label{phis}
   \>.
\end{equation}
Unlike Eq.~\eqref{lNd3.e:Vphi} we notice that spontaneous symmetry
breaking does not lead to massless particles.  This apparent defect in
the case of the Hartree approximation of the $O(N)$ model has been
discussed extensively in the literature; a review of the literature
and the solution of how to restore the Goldstone theorem in this
approximation has recently been given in Ref.~\cite{r:IRK,r:IRHK}.

When $\phi = 0$, the minimum of the potential is
\begin{align}
   V_H(\phi=0)
   =
   \frac{1}{24\pi}
   ( \, |m| - |\rho| \, )^2 \,
   ( \, |m| + 2 \, |\rho| \, )
   \>,
\end{align}
which will reach its lowest value for $m=|\rho|$, or $R=0$.  When
$\phi \neq 0$, Eq.~\eqref{phis} requires that the minimum occurs at
positive $\rho$, so that at the minimum
\begin{align}
   V_H(\phi\neq 0,\rho)
   =
   V_H(\phi=0,\rho)
   -
   \frac{\rho^2}{9} \,
   \Bigl ( \phi^2
      - \frac{2\rho}{\lambda} \,
   \Bigl ) \>.
\end{align}
From this we determine
\begin{align}
   V_H(\phi_+,\rho)
   &=
   V_H(\phi=0,\rho)
   \>,
   \\
   V_H(\phi_-,\rho)
   &=
   V_H(\phi=0,\rho)
   +
   \frac{4\rho^3}{27\lambda} \>.
\end{align}
We conclude that an absolute minimum is located at $\phi_+$, which
has the same energy as the minimum at $\phi = 0$.

When $\phi = 0$, the phase structure of the vacuum in the Hartree
approximation is the same as in the leading order large-$N$ case, with
the replacement $\lambda \rightarrow 3\lambda$.  When $\phi \neq 0$,
however, the phase structure in the Hartree approximation is
different. We have
\begin{equation}
   m
   =
   \rho
   =
   \frac{2 \mu_R}
        { ( 3 \lambda ) / (8\pi) - 1 }
   > 0 \>,
   \label{rho_H}
\end{equation}
which can be satisfied if either $\mu_R>0$ and $3\lambda
> \lambda_c /2 = 8\pi$, or $\mu_R<0$ and $\lambda < \lambda_c/2$.

So, even though the phase structure of the leading-order large N and
the Hartree approximation for $N=1$ obey the same equations when the
symmetry is not broken, for the broken-symmetry case the two theories
are quite different. The Hartree approximation yields finite masses
for the fermion and boson masses, whereas in the leading-order large N
the particles are massless in the broken-symmetry phase for
$N=1$. Furthermore the degenerate ground-state structure differs in
the two theories.  We summarize the Hartree results in
Fig.~\ref{f:phase}.

For completeness, we note here that the effective potential at
finite temperature in the Hartree approximation, is obtained as
\begin{multline}
   V_H(\phi,\rho,R;\beta)
   =
   \frac{1}{2} \, m^2 \, \phi^2
   -
   \frac{\lambda}{2} \, \rho \, \phi^4
   +
   \frac{\lambda^2}{8} \, \phi^6
   \\
   +
   \frac{2}{3\lambda} \, R \, ( \, \mu_R - \rho \, )
   -
   \frac{1}{12 \pi} \,
   \bigl ( \,
      | m |^3
      -
      | \rho |^3 \,
   \bigr )
   \\
   +
   \frac{1}{\beta}
   \int_{0}^{+\infty} \frac{k \, \rd k}{2 \pi} \,
   \ln
   \biggl [ \,
      \frac{1 - \exp ( - \beta \omega_k )}
           {1 + \exp ( - \beta \omega'_k )}
   \biggr ] \>.
\end{multline}
At the minimum of the potential, $\rho$ and $R$ satisfy the finite
temperature gap equations:
\begin{subequations}
\begin{align}
   & \rho
   =
   \mu_R
   +
   \frac{3\lambda}{4}
   \Bigl \{
      \phi^2
      \!-\!
      \frac{ | m | }{4\pi}
      \!-\!
      \frac{1}{2\pi\beta}
      \ln \bigl [ 1 - \exp(-\beta | m | ) \bigr ]
   \Bigl \}
   \\
   & R
   =
   \frac{3\lambda}{2} \rho
   \Bigl \{
      \phi^2
      - \frac{\lambda}{2} \phi^4
   \\ \notag &
      -\!
      \frac{1}{2\pi\beta}
      \Bigl \{
         \ln \bigl [ \, 2 \sinh(\beta | m |/2) \bigr ]
         \!-\!
         \ln \bigl [ \, 2 \cosh(\beta |\rho|/2) \bigr ]
      \Bigr \}
   \Bigr \} \>.
\end{align}
\end{subequations}
Finally, \emph{at the minimum}, the potential can be written as:
\begin{align}
   V_{\mathrm{eff}}& (\phi,\rho,m;\beta)
   =
   \frac{1}{2} \rho^2 \phi^2
   -
   \frac{\lambda}{2} \, \rho \, \phi^4
   +
   \frac{\lambda^2}{8} \, \phi^6
   \notag \\ &
   +
   \frac{1}{24\pi}
   \bigl ( \, | m |  - | \rho | \, \bigr )^2
   ( | m | + 2 | \rho | )
   \notag \\ &
   +
   \frac{m^2 - \rho^2}{4\pi\beta} \,
   \ln \bigl [ \, 1 - \exp(-\beta | m | ) \, \bigr ]
   \notag \\ &
   +
   \frac{1}{\beta}
   \int_{0}^{+\infty} \frac{k \, \rd k}{2 \pi} \,
   \ln
   \biggl [ \,
      \frac{1 - \exp ( - \beta \omega_k )}
           {1 + \exp ( - \beta \omega'_k )}
   \biggr ] \>.
\end{align}

%
%
\section{Conclusions}
\label{s:conclusions}

We have computed the effective potentials for a three-dimensional
supersymmetric $\Phi^4$ model in the large-$N$ and Hartree
approximations at zero temperature and at finite temperature. Both
models lead to a rich degenerate ground-state structure.  We find that
the ground state preserves supersymmetry but can have different
structure depending on the choice of coupling constant, renormalized
mass and the approximation scheme. One interesting choice of
parameters leads to the coexistence of a phase with broken and
unbroken $O(N)$ symmetry (or parity symmetry if $N=1$).  The existence
of this situation leads to the interesting question of which vacuum an
initial state prepared at high temperature will relax into. This will
be the subject of a future investigation. Another interesting question
is whether the resummed next to leading order in large-$N$
approximation obtained from the self consistent Schwinger-Dyson
equations will lift the degeneracy of the vacuum.  The main point of
this paper was to present the conceptual (and calculational) framework
for doing dynamical simulations in supersymmetric quantum field
theories.

%
%
\begin{acknowledgments}
We would like to thank the Santa Fe Institute for hospitality where
most of this work was done.
\end{acknowledgments}
%
%
\appendix
%
%
%
\section{Majorana representation in 2+1 dimensions}
\label{a.s:majorana}

In three dimensions, we can choose the Dirac $\gamma$-matrices to
satisfy a two-dimensional Clifford algebra:
\begin{equation}
   \AntiComm{\gamma^{\mu}}{\gamma^{\nu}}
   =
   \gamma^{\mu} \gamma^{\nu}
   +
   \gamma^{\nu} \gamma^{\mu}
   =
   2 \, \eta^{\mu\nu} \>,
\end{equation}
with $\eta^{\mu\nu} = \text{diag}(1,-1,-1)$.  The Majorana
representation in 2+1 dimensions is given by the choice:
\begin{align}
   \gamma^0
   &=
   \phantom{-}
   ( \, \gamma^0 \, )^{\dagger}
   =
   \phantom{i}
   \sigma^{2}
   =
   \begin{pmatrix}
      0 & -i \\
      i & 0
   \end{pmatrix}
   \>,
   \label{e:majoranagamma0} \\
   \gamma^1
   &=
   -
   ( \, \gamma^1 \, )^{\dagger}
   =
   i \sigma^{3}
   =
   \begin{pmatrix}
      i & 0 \\
      0 & -i
   \end{pmatrix}
   \>,
   \label{e:majoranagamma1}
   \\
   \gamma^2
   &=
   -
   ( \, \gamma^2 \, )^{\dagger}
   =
   i \sigma^{1}
   =
   \begin{pmatrix}
      0 & i \\
      i & 0
   \end{pmatrix}
   \>.
   \label{e:majoranagamma2}
\end{align}
With these choices, $( \, \gamma^\mu \, )^{\ast} = - \gamma^\mu$, and
$( \, \gamma^\mu \, )^{\dagger} = - ( \, \gamma^\mu \, )^{T}$, which
is just the opposite of the situation for the Weyl representation.
$\beta$, $\gamma^5$, the Hermitian conjugate ($\calC$) and complex
conjugate ($\calCstar$) operations are defined by:
\begin{alignat}{2}
   \beta \, \gamma^{\mu} \, \beta
   &=
   ( \, \gamma^\mu \, )^{\dagger}
   \>,
   &&\\
   \gamma^5 \, \gamma^{\mu} \, \gamma^5
   &=
   - \gamma^\mu \>.
   &&\\
    \calC \, \gamma^{\mu} \, \calC^{-1}
   &=
   -
   ( \, \gamma^\mu \, )^{T}
   &&=
   ( \, \gamma^\mu \, )^{\dagger}
   \\
   \calCstar \, \gamma^{\mu} \, \calCstar^{-1}
   &=
   -
   ( \, \gamma^\mu \, )^{\ast}
   &&=
   \gamma^\mu  \>.
\end{alignat}
From which we take:
\begin{align}
   \beta
   &=
   ( \, \beta \, )^{\dagger}
   =
   -
   ( \, \beta \, )^{T}
   =
   -
   ( \, \beta \, )^{\ast}
   =
   ( \, \beta \, )^{-1}
   \notag \\ &
   =
   \gamma^0
   =
   \begin{pmatrix}
      0 & -i \\
      i & 0
   \end{pmatrix}
   \>,
   \label{e:manoranabeta} \\
   \gamma^5
   &=
   ( \, \gamma^5 \, )^{\dagger}
   =
   ( \, \gamma^5 \, )^{T}
   =
   ( \, \gamma^5 \, )^{\ast}
   =
   ( \, \gamma^5 \, )^{-1}
   \notag \\ &
   =
   -
   \gamma^0 \gamma^1
   =
   \begin{pmatrix}
      0 & 1 \\
      1 & 0
   \end{pmatrix}
   \>.
   \label{e:majoranagamma5}
   \\
   \calC
   &=
   ( \, \calC \, )^{\dagger}
   =
   -
   ( \, \calC \, )^{T}
   =
   -
   ( \, \calC \, )^{\ast}
   =
   ( \, \calC \, )^{-1}
   \notag \\ &
   =
   \beta
   =
   \begin{pmatrix}
      0 & -i \\
      i & 0
   \end{pmatrix}
   \>,
   \label{e:majoranacalC} \\
   \calCstar
   &=
   ( \, \calCstar \, )^{\dagger}
   =
   ( \, \calCstar \, )^{T}
   =
   ( \, \calCstar \, )^{\ast}
   =
   ( \, \calCstar \, )^{-1}
   \notag \\ &
   =
   \calC \beta
   =
   \begin{pmatrix}
      1 & 0 \\
      0 & 1
   \end{pmatrix}
   \>.
   \label{e:majoranacalCstar}
\end{align}
With these selections, $\gamma^5$, $\beta$, and $\calC$ obey the
relations:
\begin{equation}
   \AntiComm{\gamma^5}{\beta}
   =
   \Comm{\beta}{\calC}
   =
   \AntiComm{\calC}{\gamma^5}
   =
   0 \>.
\end{equation}
These commutation and anticommutation relations differ from the
situation in 3+1 dimensions.  We write a two-component spinor $\psi$
and $\bar{\psi} = \psi^{\dagger} \beta$ as:
\begin{equation}
   \psi
   =
   \begin{pmatrix} \psi_1 \\ \psi_2 \end{pmatrix} \>,
   \qquad
   \bar{\psi}
   =
   \begin{pmatrix}
      i \psi_2^{\ast} , & - i \psi_1^{\ast}
   \end{pmatrix} \>.
\end{equation}
Then a Majorana spinor satisfies:
\begin{equation}
   \psi
   =
   \psi_c
   =
   \calCstar \, \psi^{\ast}
   =
   \psi^{\ast} \>,
\end{equation}
which means that Majorana spinors are \emph{real}, $\psi_1^{\ast} =
\psi_1$ and $\psi_2^{\ast} = \psi_2$.

%
%
%
\section{Majorana Grassmann quantities}
\label{a.s:grassmann}

In the Majorana representation, we define a \emph{real}
two-component column Majorana Grassmann spinor $\theta$ by:
\begin{equation}\label{e:MG.thetadef}
   \theta
   =
   \begin{pmatrix}
      \theta_1 \\
      \theta_2
   \end{pmatrix} \>,
\end{equation}
with $\theta_1$ and $\theta_2$ real.  The \emph{imaginary}
two-component row spinor $\bar{\theta}$ is defined by:
\begin{equation}\label{e:MG.barthetadef}
   \bar{\theta}
   =
   \theta^{T} \, \beta
   =
   \begin{pmatrix}
      i \theta_2, &
      - i \theta_1
   \end{pmatrix}
   \equiv
   \begin{pmatrix}
      \bar{\theta}_1, &
      \bar{\theta}_2
   \end{pmatrix} \>.
\end{equation}
where $\bar{\theta}_1 = i \theta_2$ and $\bar{\theta}_2 = - i
\theta_1$ are \emph{imaginary} Grassmann variables.  In component
notation, we have:
\begin{equation}\label{e:MG.thetabartheta}
   \bar{\theta}_b
   =
   \theta_{a} \, \beta_{ab} \>,
   \qquad \text{and} \qquad
   \theta_b
   =
   \bar{\theta}_{a} \, \beta_{ab} \>.
\end{equation}
The Grassmann variables all anticommute:
\begin{equation}\label{e:MG.anticomms}
   \AntiComm{\theta_{a}}{\theta_{b}}
   =
   \AntiComm{\theta_{a}}{\bar{\theta}_{b}}
   =
   \AntiComm{\bar{\theta}_{a}}{\theta_{b}}
   =
   \AntiComm{\bar{\theta}_{a}}{\bar{\theta}_{b}}
   =
   0 \>.
\end{equation}
This means that $\theta_a^2 = 0$ and $\bar{\theta}_a^2 = 0$ (no sum
over $a$ required here).  We find the useful relations:
\begin{equation}\label{e:thetatheta}
   \bar{\theta} \cdot \theta
   =
   i \, ( \, \theta_2 \theta_1 - \theta_1 \theta_2 \, )
   =
   - 2 i \, \theta_1 \theta_2 \>,
   \
   \bar{\theta}_{a} \theta_{b}
   =
   \frac{1}{2} \, \bar{\theta} \cdot \theta \>
   \delta_{ab} \>.
\end{equation}
Grassmann derivatives operators are defined by:
\begin{equation}\label{e:MG.partialsdefs}
   \partial_{a}
   =
   \frac{\partial}{\partial \theta_{a}} \>,
   \qquad \text{and} \quad
   \bar{\partial}_{a}
   =
   \frac{\partial}{\partial \bar{\theta}_{a}} \>,
\end{equation}
so that $\partial_1$ and $\partial_2$ are \emph{real} and
$\bar{\partial}_1 = - i \partial_2$ and $\bar{\partial}_2 = i
\partial_1$ are \emph{imaginary}.  We follow convention and reverse
the definition of row and column matrices for the derivatives, and
write:
\begin{equation}\label{e:MG.matpartialsdefs}
   \bar{\partial}
   \equiv
   \begin{pmatrix}
      \bar{\partial}_1 \\
      \bar{\partial}_2
   \end{pmatrix} \>,
   \qquad \text{and} \quad
   \partial
   \equiv
   \begin{pmatrix}
      \partial_1 &
      \partial_2
   \end{pmatrix}
   =
   \bar{\partial}^{T} \, \beta \>.
\end{equation}
In component notation, we have:
\begin{equation}\label{e:MG.partialbarpartial}
   \partial_{b}
   =
   \bar{\partial}_{a} \, \beta_{ab} \>,
   \qquad \text{and} \quad
   \bar{\partial}_b
   =
   \partial_a \,\beta_{ab} \>.
\end{equation}
So we find:
\begin{gather}\label{e:MG.partialsthetas}
   \partial_a \theta_b
   =
   \bar{\partial}_a \bar{\theta}_b
   =
   \delta_{ab} \>,
   \\
   \partial_a \bar{\theta}_b
   =
   \beta_{ab} \>,
   \qquad
   \bar{\partial}_a \theta_b
   =
   \beta_{ba} \>,
   \notag
\end{gather}
so that $\bar{\theta}$ is not independent of $\theta$.  The
differential operators obey the following anticommutator relations:
\begin{equation}\label{e:MG.derivsanticomms}
   \AntiComm{\partial_a}{\partial_b}
   =
   \AntiComm{\bar{\partial}_a}{\bar{\partial}_b}
   =
   \AntiComm{\partial_a}{\bar{\partial}_b}
   = 0 \>,
\end{equation}
This means that $\partial_a^2 = 0$ and $\bar{\partial}_a^2 = 0$ (no
sum over $a$ required here).  We also have:
\begin{gather}\label{e:MG.partialthetatheta}
   \AntiComm{\partial_a}{\theta_b}
   =
   \AntiComm{\bar{\partial}_a}{\bar{\theta}_b}
   =
   \delta_{ab}
   \\
   \AntiComm{\partial_a}{\bar{\theta}_b}
   =
   \beta_{ab} \, \>,
   \qquad
   \AntiComm{\bar{\partial}_a}{\theta_b}
   =
   \beta_{ba} \, \>.
   \notag
\end{gather}
It is useful to define the integration measure with a factor of
$i/2$ so that we have the relations:
\begin{gather}\label{e:thetaidentities}
   \rd^2 \theta
   =
   \frac{1}{2} \, \rd \bar{\theta}_1 \, \rd \theta_1
   =
   \frac{i}{2} \, \rd \theta_2 \, \rd \theta_1 \>,
   \\ \notag
   \int \rd^2 \theta
   =
   \int \rd^2 \theta \> \theta_a
   =
   0 \>,
   \\ \notag
   \int \rd^2 \theta \> \bar{\theta}_{a} \, \theta_{b}
   =
   \frac{1}{2} \, \delta_{ab} \>,
   \quad
   \int \rd^2 \theta \> \bar{\theta} \cdot \theta
   =
   1  \>.
\end{gather}
With this convention, the Grassmann two-dimensional delta function
for two anticommuting Grassmann quantities is given by:
\begin{align}\label{mg.e:deltafunctiondef}
   \delta^2(\theta - \theta')
   & =
   ( \, \bar{\theta} - \bar{\theta}' \, ) \cdot
   ( \, \theta - \theta' \, )
   \notag \\
   &=
   \bar{\theta} \cdot \theta
   -
   \bar{\theta} \cdot \theta'
   -
   \bar{\theta}' \cdot \theta
   +
   \bar{\theta}' \cdot \theta'
   \notag \\
   &=
   \bar{\theta} \cdot \theta
   -
   2 \, \bar{\theta} \cdot \theta'
   +
   \bar{\theta}' \cdot \theta'  \>.
\end{align}
If $\theta$ and $\phi$ are two anticommuting Majorana Grassmann
spinors, we have the useful identities:
\begin{gather}
   \bar{\theta} \cdot \phi
   =
   \bar{\phi} \cdot \theta \>,
   \qquad
   \bar{\theta} \cdot \gamma^{\mu} \cdot \phi
   =
   -
   \bar{\phi} \cdot \gamma^{\mu} \cdot \theta \>,
   \\
   \bar{\theta}_a \, \theta_b
   =
   \delta_{a,b} \, ( \bar{\theta} \cdot \theta ) / 2 \>,
   \qquad
   \bar{\theta} \cdot \gamma^{\mu} \cdot \theta
   =
   0 \>,
   \\
   \partial_a ( \bar{\theta} \cdot \theta )
   =
   ( \partial_a \bar{\theta}_b ) \, \theta_b
   -
   \bar{\theta}_b \, ( \partial_a \bar{\theta}_b )
   =
   - 2 \, \bar{\theta}_a \>,
   \\
   \bar{\partial}_a ( \bar{\theta} \cdot \theta )
   =
   ( \bar{\partial}_a \bar{\theta}_b ) \, \theta_b
   -
   \bar{\theta}_b \, ( \bar{\partial}_a \bar{\theta}_b )
   =
   + 2 \, \theta_a \>,
   \\
   ( \partial \cdot \bar{\partial} ) \,
   ( \bar{\theta} \cdot \theta )
   =
   4 \>.
\end{gather}
The supercharge generators $Q$ and $\bar{Q}$ and the superderivative
operators $D$ and $\bar{D}$ are defined by:
\begin{alignat}{2}
   Q
   &=
   -
   \bar{\partial}
   -
   i \, \Slash{\partial} \cdot \theta  \>,
   & \qquad
   D
   &=
   +
   \bar{\partial}
   -
   i \, \Slash{\partial} \cdot \theta \>,
   \label{mg.e:QDdefs} \\
   \bar{Q}
   &=
   +
   \partial
   +
   i \, \bar{\theta} \cdot \Slash{\partial} \>,
   & \qquad
   \bar{D}
   &=
   -
   \partial
   +
   i \, \bar{\theta} \cdot  \Slash{\partial} \>.
   \notag
\end{alignat}
where we have used the Dirac slash notation, $\Slash{\partial} =
\gamma^{\mu} \partial_{\mu}$.  The $Q$ operators obey the
superalgebra:
\begin{align}\label{e:Qsuperalgebra}
   \AntiComm{Q_{a}}{\bar{Q}_{b}}
   =
   -
   2 i \, ( \,  \gamma^{\mu} \, )_{ab} \, \partial_{\mu} \>,
   \\ \notag
   \qquad
   \AntiComm{Q_{a}}{Q_{b}}
   =
   0 \>, \qquad
   \AntiComm{\bar{Q}_{a}}{\bar{Q}_{b}}
   =
   0 \>.
\end{align}
Thus we find:
\begin{align}
   Q \cdot Q
   =
   \bar{Q} \cdot \bar{Q}
   =
   0 \>,
   \\ \notag
   Q \cdot \bar{Q} + \bar{Q} \cdot Q
   =
   - 2 i \, \mathrm{Tr} \{ \gamma^{\mu} \} \,  \partial_{\mu}
   =
   0 \>.
\end{align}
The $D$ operators satisfy a similar superalgebra but with a reversed
sign:
\begin{align}\label{e:Dsuperalgebra}
   \AntiComm{D_{a}}{\bar{D}_{b}}
   =
   2 i \, ( \,  \gamma^{\mu} \, )_{ab} \, \partial_{\mu} \>.
   \\ \notag
   \AntiComm{D_{a}}{D_{b}}
   =
   0 \>, \quad
   \AntiComm{\bar{D}_{a}}{\bar{D}_{b}}
   =
   0 \>,
\end{align}
So we find:
\begin{align}\label{e:a.DbarDanticom}
   D \cdot D
   =
   \bar{D} \cdot \bar{D}
   =
   0 \>,
   \\ \notag
   D \cdot \bar{D} + \bar{D} \cdot D
   =
   2 i \, \mathrm{Tr} \{ \gamma^{\mu} \} \,  \partial_{\mu}
   =
   0 \>.
\end{align}
The $Q$ and $D$ operators anticommute:
\begin{align}\label{e:DQsuperalgebra}
   \AntiComm{D_{a}}{Q_{b}}
   & =
   \AntiComm{\bar{D}_{a}}{\bar{Q}_{b}}
   \\ \notag
   & =
   \AntiComm{D_{a}}{\bar{Q}_{b}}
   =
   \AntiComm{\bar{D}_{a}}{Q_{b}}
   =
   0 \>.
\end{align}
The operators $Q$ and $D$ are related by:
\begin{equation}\label{e:a.QDrelation}
   Q = D - 2 \bar{\partial}
   \>, \qquad
   \bar{Q} = \bar{D} + 2 \partial \>.
\end{equation}
We also find:
\begin{align}\label{e:a:barDD}
   \bar{D} \cdot D
   &=
   \bigl ( \,
      -
      \partial
      +
      i \, \bar{\theta} \cdot \gamma^{\mu} \, \partial_{\mu} \,
   \bigr ) \cdot
   \bigl ( \,
      \bar{\partial}
      -
      i \, \gamma^{\nu} \cdot \theta \, \partial_{\nu} \,
   \bigr )
   \notag \\
   &=
   -
   \partial \cdot \bar{\partial}
   -
   i \,
   \bigl ( \,
      \partial \cdot \gamma^{\mu} \cdot \theta
      -
      \bar{\theta} \cdot \gamma^{\mu} \cdot \bar{\partial} \,
   \bigr ) \, \partial_{\mu}
   \notag \\ & \qquad
   +
   \bar{\theta} \cdot \gamma^{\mu} \cdot \gamma^{\nu} \cdot \theta \,
   \partial_{\mu} \, \partial_{\nu}
   \\ \notag
   &=
   -
   \partial \cdot \bar{\partial}
   +
   2 i \, \bar{\theta} \cdot \gamma^{\mu} \cdot \bar{\partial} \,
   \partial_{\mu}
   +
   ( \bar{\theta} \cdot \theta ) \, \square  \>,
\end{align}
where $\partial \cdot \bar{\partial} = \partial_1 \,
\bar{\partial}_1 + \partial_2 \, \bar{\partial}_2$ and $\square =
\partial_t^2 -
\partial_x^2$.  From Eq.~\eqref{mg.e:deltafunctiondef}, we find:
\begin{equation}
   \bar{D} \cdot D \, \delta(\theta,\theta')
   =
   - 4 \,
   e^{  i \bar{\theta} \cdot \Slash{\partial} \cdot \theta' } \>,
\end{equation}

%
%
\section{Temperature dependent supergreen functions}
\label{tg.s:tempgreen}

We use a complex time formalism for the temperature dependent Green
functions.  The superperiodic boundary condition on the superfield
is then given by:
\begin{equation}\label{tg.e:superthermalboundry}
   \Phi_i(\mathbf{x},\beta;\theta)
   =
   \Phi_i(\mathbf{x},0;-\theta)  \>.
\end{equation}
So the fields and Green functions can be expanded in a Fourier
series for the imaginary time variable, and a Fourier integral for
the space variable.  We consider here only spacial homogeneous
systems.  Thus for the fields, we write:
\begin{align}\label{tg.e:momspace}
   & \Phi(\mathbf{x},\tau;\theta)
   =
   \int \frac{\rd^{2} k}{(2\pi)^{2}} \,
   \frac{e^{i\mathbf{k}\cdot\mathbf{x}}}{\beta} \sum_{n=-\infty}^{+\infty}
   \biggl \{ \,
      \tilde{\phi}_n(k) \,  e^{i \omega_n \tau}
   \notag \\ &
      +
      \bar{\theta} \cdot \tilde{\psi}_n(k) \,
      e^{i \omega'_n \tau}
      +
      \frac{1}{2} \, ( \bar\theta \cdot \theta ) \,
      \tilde{F}_n(k) \, e^{i \omega_n \tau} \,
   \biggr \} \>,
\end{align}
where $\beta = 1 / k_{\text{B}} T$ and $\omega_n$ and $\omega'_n$
are the Bose and Fermi Matsubara frequencies:
\begin{equation}\label{tg.e:matsubara}
\begin{split}
   \omega_n
   &=
   \pi \, 2 n / \beta \>, \qquad
   \text{for boson fields.}
   \\
   \omega'_n
   &=
   \pi \, ( 2 n  + 1 ) / \beta \>, \qquad
   \text{for fermi fields.}
\end{split}
\end{equation}
The Green functions are expanded according to:
\begin{align}\label{tg.e:Greenexpnasion}
   G(x,\tau,\theta; & x',\tau',\theta')
   =
   \int \frac{\rd^{2} k}{(2\pi)^{2}} \,
   \frac{1}{\beta}
   \\ \notag & \times
   \sum_{n=-\infty}^{+\infty}
   \tilde{G}_n(\mathbf{k};\theta,\theta') \,
   e^{i
      [ \,
         \mathbf{k}\cdot (\mathbf{x} - \mathbf{x}')
         +
         \omega_n ( \tau - \tau') \,
      ] }  \>.
\end{align}
$\tilde{G}_n(\mathbf{k};\theta,\theta')$ satisfies:
\begin{align}
   & \Bigl [
      -
      \partial \cdot \bar\partial
      +
      2 i \, \bar{\theta} \cdot
      \kslash'_n
      \cdot \bar{\partial}
      +
      \bar{\theta} \cdot \theta \, k_n^2
      +
      \chi(\theta)
   \Bigr ] \tilde{G}_n(\mathbf{k};\theta,\theta')
   \notag \\ & = \
   \delta^2(\theta - \theta')  \>.
\label{tg.e:tildeGdiffeq}
\end{align}
Here $k_n^{\mu} = (\omega_n,\mathbf{k})$ and $k'_n{}^{\mu} =
(\omega'_n,\mathbf{k})$ are the Euclidean vectors, with $\kslash =
\gamma^{\mu} ( k_n )_{\mu}$ and $\kslash' = \gamma^{\mu} ( k'_n
)_{\mu}$.  We find:
\begin{align}
   &
   \tilde{G}_n(\mathbf{k};\theta,\theta')
   =
   \bar\theta \cdot
      \frac{ i \kslash'_n - \rho }{ k'_n{}^2 + \rho^2 } \cdot
   \theta'
   \\ \notag &
   + \frac{
      1
      +
      \tfrac{1}{2} \,
      ( \,
         \bar\theta \cdot \theta
         +
         \bar\theta' \cdot \theta' \,
      ) \, \rho
      -
      \tfrac{1}{4} \,
      \bar\theta \cdot \theta \, \bar\theta' \cdot \theta' \,
      ( \,  k_n^2 + \rho^2 \, ) \,
        }{ k_n^2 + \rho^2 + R }
   \>,
\end{align}
in agreement with MZJ \cite{ref:MMZJ}[Eq.~(2.7)].  The diagonal
elements of $\tilde{G}_n(\mathbf{k};\theta,\theta')$ are given by:
\begin{equation}\label{tg.e:Geffpotdiag}
   \tilde{G}_n(\mathbf{k};\theta,\theta)
   =
   \frac{
      1
      +
      \bar\theta \cdot \theta \, \rho
        }{ k_n^2 + \rho^2 + R }
   -
   \frac{ \bar\theta \cdot \theta \, \rho }{ k'_n{}^2 + \rho^2 } \>.
\end{equation}
We define:
\begin{align}
   \omega_k
   =
   \sqrt{ k^2 + \rho^2 + R } \>,
   \label{tg.e:defomegak} \\
   \omega'_k
   =
   \sqrt{ k^2 + \rho^2 } \>.
   \label{tg.e:defomegapk}
\end{align}
Then the sum over $n$ is given by:
\begin{equation}\label{tg.e:sums}
\begin{split}
   \sum_{n=-\infty}^{+\infty}
   \frac{1}{\omega_n^2 + \omega_k^2}
   & =
   \frac{\beta}{2 \, \omega_k} \, \coth(\beta\omega_k /2)
   \\ & =
   \frac{\beta}{2 \, \omega_k} \,
   \Bigl [ \,
      1 + 2 \, n_{-}(\beta\omega_k) \,
   \Bigr ] \>,
   \\
   \sum_{n=-\infty}^{+\infty} \frac{1}{\omega'_n{}^2 + \omega'_k{}^2}
   & =
   \frac{\beta}{2 \, \omega'_k} \, \tanh(\beta\omega'_k /2)
   \\ & =
   \frac{\beta}{2 \, \omega'_k} \,
   \Bigl [ \,
      1 - 2 \, n_{+}(\beta\omega'_k) \,
   \Bigr ] \>,
\end{split}
\end{equation}
where we have $n_{\pm}(x) = [ {e^{x} \pm 1} ]^{-1}$. So
\begin{align}
   \frac{1}{\beta} \sum_{n=-\infty}^{+\infty}
   & \tilde{G}_n(\mathbf{k};\theta,\theta)
   =
   \frac{\coth(\beta\omega_k /2)}{2 \omega_k}
   \\ \notag & +
   \bar\theta \cdot \theta \,
   \biggl [ \,
      \frac{\rho \coth(\beta\omega_k /2)}{2 \omega_k}
      -
      \frac{\rho \tanh(\beta\omega'_k /2)}{2 \omega'_k} \,
   \biggr ]
\end{align}
For the large-$N$ approximation, the thermal gap equation becomes:
\begin{align}
   &
   \chi(\theta)
   \!=\!
   2 \mu
   \!+\!
   \frac{\lambda}{2} \sum_i \Phi_i^2(\theta)
   \!+\!
   \frac{\lambda}{2}
   \int \frac{\rd^{2} k}{(2\pi)^{2}}
   \biggl \{
      \frac{\coth(\beta\omega_k /2)}{2 \omega_k}
   \notag \\ &
\label{tg.e:superthermalgapii}
      +
      \bar\theta \cdot \theta
      \biggl [
         \frac{\rho \coth(\beta\omega_k /2)}{2 \omega_k}
         -
         \frac{\rho \tanh(\beta\omega'_k /2)}{2 \omega'_k}
      \biggr ]
   \biggr \}
\end{align}
from which we find, for $N=1$:
\begin{subequations}\label{tg.e:thermalgapcomp}
\begin{align}
   \rho
   &=
   \mu
   +
   \frac{\lambda}{4} \phi^2
   +
   \frac{\lambda}{4}
   \int^{\Lambda} \frac{\rd^{2} k}{(2\pi)^{2}}
   \frac{\coth(\beta\omega_k /2)}{2 \omega_k} \>,
   \label{tg.e:tgci} \\
   R
   &=
   \frac{\lambda}{2} \, \phi \, F
  \label{tg.e:tgcii}
  \\ \notag &
   +
   \frac{\lambda}{2}
   \int \frac{\rd^{2} k}{(2\pi)^{2}} \,
      \biggl [ \,
         \frac{\rho \coth(\beta\omega_k /2)}{2 \omega_k}
         -
         \frac{\rho \tanh(\beta\omega'_k /2)}{2 \omega'_k} \,
      \biggr ]
\end{align}
\end{subequations}
Here we have used a finite cutoff $\Lambda$ so as to make the integral
in Eq.~\eqref{tg.e:tgci} finite.  We renormalize this by defining:
\begin{equation}\label{tgd3.e:muR}
   \mu_R
   =
   \mu
   +
   \frac{\lambda}{4}
   \int_{0}^{\Lambda} \frac{k \, \rd k}{2\pi} \,
      \frac{1}{ 2 \sqrt{k^2} } \>.
\end{equation}
so that Eq.~\eqref{tg.e:tgci} becomes:
\begin{equation}\label{tgd3.e:trho}
\begin{split}
   \rho
   &=
   \mu_R
   +
   \frac{\lambda}{4} \, \phi^2
   -
   \frac{\lambda}{8\pi \beta} \,
   \ln \bigl [ \, 2 \sinh ( \beta m / 2 ) \bigr ] \>,
\end{split}
\end{equation}
where again $m^2 = \rho^2 + R$.  Eq.~\eqref{tg.e:tgcii} is finite,
and becomes:
\begin{equation}\label{tgd3.e:tR}
\begin{split}
   & R
   =
   \frac{\lambda}{2} \, \phi \, F
   \\ \notag &
   -
   \frac{\lambda \rho}{4\pi\beta} \,
   \Bigl \{ \,
      \ln \bigl [ 2 \sinh(\beta m / 2) \bigr ]
      -
      \ln \bigl [ 2 \cosh(\beta \rho / 2) \bigr ] \,
   \Bigr \} \>.
\end{split}
\end{equation}
Multiplying \eqref{tgd3.e:trho} by $2 \rho$ and subtracting it from
\eqref{tgd3.e:tR} yields:
\begin{equation}\label{tgd3.e:mtoR}
   R
   =
   \rho \,
   \Bigl \{ \,
      2 \, ( \rho - \mu_R )
      -
      \frac{\lambda}{4\pi\beta} \,
      \ln \bigl [ 2 \cosh(\beta \rho / 2) \bigr ] \,
   \Bigr \} \>,
\end{equation}
along the line where $F = \rho \, \phi$.

The effective potential at finite temperature for large-$N$ is
written as the sum of two terms:
\begin{equation}\label{tg.e:VeffVcVq}
   V_{\text{eff}}(\phi,F;\rho,R;\beta)
   =
   V_{\text{c}}(\phi,F;\rho,R)
   +
   V_{\text{q}}(\rho,R;\beta) \>,
\end{equation}
where the classical part is given by:
\begin{equation}\label{tg.e:Vci}
   V_{\text{c}}(\phi,F;\rho,R)
   =
   \rho \, \phi \, F
   -
   \frac{1}{2} \, F^2
   +
   \frac{1}{2} \, R \, \phi^2
   +
   \frac{2}{\lambda} \, R \, ( \, \mu_R - \rho \, ) \>.
\end{equation}
At the minimum, where $F = \rho \, \phi$,
$V_{\text{c}}(\phi,\rho\,\phi;\rho,R)$ becomes:
\begin{equation}\label{tg.e:Vcii}
   V_{\text{c}}(\phi;\rho,R)
   =
   \frac{1}{2} \, m^2 \, \phi^2
   +
   \frac{2}{\lambda} \, R \, ( \, \mu_R - \rho \, ) \>.
\end{equation}
For the quantum part $V_{\text{q}}(\rho,R)$, we define
$W_{\text{q}}(\rho,R;\theta)$ by:
\begin{equation}\label{tg.e:VqW}
   V_{\text{q}}(\rho,R;\beta)
   =
   \int \rd^2 \theta \, W_{\text{q}}(\rho,R;\theta;\beta) \>.
\end{equation}
For $d=3$, $W_{\text{q}}(\rho,R;\theta)$ is given by:
\begin{align}\label{tg3d.e:Wdef}
   W_{\text{q}}(\rho,R; & \theta;\beta)
   =
   \frac{1}{8\pi} \,
   \int_{0}^{\infty} k \, \rd k
   \\ \notag & \times
   \biggl \{ \,
      \frac{1}{i\beta} \,
      \sum_{n=-\infty}^{+\infty}
      \ln \bigl [ \, \tilde{G}_n^{-1}(k;\theta,\theta) \, \bigr ]
      -
      \frac{\chi(\theta)}{2\sqrt{k^2}} \,
   \biggr \} \>.
\end{align}
Again, we can write 
\begin{align}
   \frac{\delta W_{\text{q}}(\rho,R;\theta;\beta)}
        {\delta \chi(\theta)}
\label{tg3d.e:Wqderv}
   = &
   W_0(\rho,R;\beta)
   +
   \frac{1}{2} \, \bar\theta \cdot \theta \,
   W_1(\rho,R;\beta) \>,
\end{align}
with
\begin{equation}\label{tg3d.e:W0}
   W_0(\rho,R;\beta)
   =
   -
   \frac{1}{4\pi\beta}
   \ln \bigl [ \, 2 \sinh ( \beta m / 2 ) \, \bigr ] \>,
\end{equation}
and
\begin{align}\label{tg3d.e:W1}
   W_1(\rho,R;\beta)
   =
   - \frac{N rho}{2\pi \beta}
   \Bigl \{ \,
   &   \ln \bigl [ \, 2 \sinh(\beta m / 2) \, \bigr ]
   \\ \notag & \quad
      -
      \ln \bigl [ \, 2 \cosh(\beta \rho / 2) \, \bigr ] \,
   \Bigr \} \>.
\end{align}
Again, we have
\begin{equation}
   \delta \chi(\rho,R)
   =
   2 \delta \rho
   +
   \bar\theta \cdot \theta \, \delta R \>,
\end{equation}
so
\begin{equation}
\begin{split}
   & \delta W_{\text{q}}(\rho,R;\theta;\beta)
   =
   2 \, W_0(\rho,R) \, \delta \rho
   \\ &
   +
   \bar\theta \cdot \theta \,
   \bigl [ \,
       W_0(\rho,R;\beta) \, \delta R
       +
       W_1(\rho,R;\beta) \, \delta \rho  \,
   \bigr ]
\end{split}
\end{equation}
We now want to find a common function $V_{\text{q}}(\rho,R;\beta)$
such that:
\begin{equation}
\begin{split}
   \frac{\partial V_{\text{q}}(\rho,R;\beta)}{\partial R}
   & =
   W_0(\rho,R;\beta) \>,
   \\
   \frac{\partial V_{\text{q}}(\rho,R;\beta)}{\partial \rho}
   & =
   W_1(\rho,R;\beta) \>.
\end{split}
\end{equation}
We can use the following identities:
\begin{equation*}
\begin{split}
   \frac{\rho \, n_{-}( \beta \omega_k)}{\omega_k}
   &=
   \frac{1}{\beta}
   \frac{\partial}{\partial \rho}
   \Bigl \{
      \ln \bigl [ 1 - \exp( - \beta \omega_k ) \bigr ]
   \Bigr \} \>,
   \\
   \frac{\rho \, n_{+}( \beta \omega'_k)}{\omega'_k}
   &=
   - \frac{1}{\beta}
   \frac{\partial}{\partial \rho}
   \Bigl \{
      \ln \bigl [ 1 + \exp( - \beta \omega'_k ) \bigr ]
   \Bigr \}  \>,
   \\
   \frac{n_{-}( \beta \omega_k)}{2 \omega_k}
   &=
   \frac{1}{\beta}
   \frac{\partial}{\partial R}
   \Bigl \{
      \ln \bigl [ 1 - \exp( - \beta \omega_k ) \bigr ]
   \Bigr \} \>.
\end{split}
\end{equation*}
So the function we seek is:
\begin{align}\label{3d.e:Veffp}
   V_q(\rho,R; & \beta)
   =
   \frac{1}{12\pi}
   \bigl (  | \rho |^3 - | m |^3  \bigr )
   \\ \notag &
   +
   \frac{1}{2\pi \beta}
   \int_{0}^{+\infty}  k  \rd k \,
      \ln \bigl [ 1 - \exp( - \beta \omega_k ) \bigr ]
   \\ \notag &
   -
   \frac{1}{2\pi \beta}
   \int_{0}^{+\infty}  k  \rd k \,
      \ln \bigl [ 1 + \exp( - \beta \omega'_k ) \bigr ]
   \>.
\end{align}
Thus the effective potential at finite temperature and along the line
$F= \rho \, \phi$ is given by:
\begin{align}\label{a.tg3d.e:Veff}
   V_{\mathrm{eff}}(\phi,\rho,R; & \beta)
   =
   \frac{1}{2} \, m^2 \, \phi^2
   +
   \frac{2}{\lambda} \, R \, ( \, \mu_R - \rho \, )
   \\ \notag &
   +
   \frac{1}{12\pi}
   \bigl ( \, | \rho |^3 - | m |^3 \, \bigr )
   \\ \notag &
   +
   \frac{1}{2\pi \beta}
   \int_{0}^{+\infty}  k \, \rd k \,
      \ln \bigl [ 1 - \exp( - \beta \omega_k ) \bigr ] \,
   \\ \notag &
   -
   \frac{1}{2\pi \beta}
   \int_{0}^{+\infty}  k \, \rd k \,
      \ln \bigl [ 1 + \exp( - \beta \omega'_k ) \bigr ] \,
   \>.
\end{align}
This expression agrees with Eq.~\eqref{lNd3.e:VNi} in the vacuum
when $\beta \rightarrow \infty$.

%
%
%
\bibliography{johns}

\begin{thebibliography}{26}
\expandafter\ifx\csname natexlab\endcsname\relax\def\natexlab#1{#1}\fi
\expandafter\ifx\csname bibnamefont\endcsname\relax
  \def\bibnamefont#1{#1}\fi
\expandafter\ifx\csname bibfnamefont\endcsname\relax
  \def\bibfnamefont#1{#1}\fi
\expandafter\ifx\csname citenamefont\endcsname\relax
  \def\citenamefont#1{#1}\fi
\expandafter\ifx\csname url\endcsname\relax
  \def\url#1{\texttt{#1}}\fi
\expandafter\ifx\csname urlprefix\endcsname\relax\def\urlprefix{URL }\fi
\providecommand{\bibinfo}[2]{#2}
\providecommand{\eprint}[2][]{\url{#2}}

\bibitem[{\citenamefont{Wess and Bagger}(1992)}]{ref:WessBagger92}
\bibinfo{author}{\bibfnamefont{J.}~\bibnamefont{Wess}} \bibnamefont{and}
  \bibinfo{author}{\bibfnamefont{J.}~\bibnamefont{Bagger}},
  \emph{\bibinfo{title}{{S}upersymmetry and Supergravity}}
  (\bibinfo{publisher}{{Princeton University Press}},
  \bibinfo{address}{Princeton, NJ}, \bibinfo{year}{1992}),
  \bibinfo{edition}{2nd} ed.

\bibitem[{\citenamefont{Cornwall et~al.}(1974)\citenamefont{Cornwall, Jackiw,
  and Tomboulis}}]{r:CJT}
\bibinfo{author}{\bibfnamefont{J.~M.} \bibnamefont{Cornwall}},
  \bibinfo{author}{\bibfnamefont{R.}~\bibnamefont{Jackiw}}, \bibnamefont{and}
  \bibinfo{author}{\bibfnamefont{E.}~\bibnamefont{Tomboulis}},
  \bibinfo{journal}{Phys. Rev. D} \textbf{\bibinfo{volume}{10}},
  \bibinfo{pages}{2428} (\bibinfo{year}{1974}).

\bibitem[{\citenamefont{Luttinger and Ward}(1960)}]{r:LW}
\bibinfo{author}{\bibfnamefont{J.~M.} \bibnamefont{Luttinger}}
  \bibnamefont{and} \bibinfo{author}{\bibfnamefont{J.~C.} \bibnamefont{Ward}},
  \bibinfo{journal}{Phys. Rev.} \textbf{\bibinfo{volume}{118}},
  \bibinfo{pages}{1417} (\bibinfo{year}{1960}).

\bibitem[{\citenamefont{Baym}(1962)}]{r:Baym62}
\bibinfo{author}{\bibfnamefont{G.}~\bibnamefont{Baym}}, \bibinfo{journal}{Phys.
  Rev.} \textbf{\bibinfo{volume}{127}}, \bibinfo{pages}{1391}
  (\bibinfo{year}{1962}).

\bibitem[{\citenamefont{Cooper et~al.}(2003{\natexlab{a}})\citenamefont{Cooper,
  Dawson, and Mihaila}}]{r:CDM02ii}
\bibinfo{author}{\bibfnamefont{F.}~\bibnamefont{Cooper}},
  \bibinfo{author}{\bibfnamefont{J.~F.} \bibnamefont{Dawson}},
  \bibnamefont{and} \bibinfo{author}{\bibfnamefont{B.}~\bibnamefont{Mihaila}},
  \bibinfo{journal}{Phys. Rev. D} \textbf{\bibinfo{volume}{67}},
  \bibinfo{pages}{056003} (\bibinfo{year}{2003}{\natexlab{a}}).

\bibitem[{\citenamefont{Cooper et~al.}(2003{\natexlab{b}})\citenamefont{Cooper,
  Dawson, and Mihaila}}]{r:CDM02}
\bibinfo{author}{\bibfnamefont{F.}~\bibnamefont{Cooper}},
  \bibinfo{author}{\bibfnamefont{J.~F.} \bibnamefont{Dawson}},
  \bibnamefont{and} \bibinfo{author}{\bibfnamefont{B.}~\bibnamefont{Mihaila}},
  \bibinfo{journal}{Phys. Rev. D} \textbf{\bibinfo{volume}{67}},
  \bibinfo{pages}{051901(R)} (\bibinfo{year}{2003}{\natexlab{b}}).

\bibitem[{\citenamefont{Aarts et~al.}(2002)\citenamefont{Aarts, Ahrensmeier,
  Baier, Berges, and Serreau}}]{r:AABBS}
\bibinfo{author}{\bibfnamefont{G.}~\bibnamefont{Aarts}},
  \bibinfo{author}{\bibfnamefont{D.}~\bibnamefont{Ahrensmeier}},
  \bibinfo{author}{\bibfnamefont{R.}~\bibnamefont{Baier}},
  \bibinfo{author}{\bibfnamefont{J.}~\bibnamefont{Berges}}, \bibnamefont{and}
  \bibinfo{author}{\bibfnamefont{J.}~\bibnamefont{Serreau}},
  \bibinfo{journal}{Phys. Rev. D} \textbf{\bibinfo{volume}{66}},
  \bibinfo{pages}{045008} (\bibinfo{year}{2002}).

\bibitem[{\citenamefont{Berges and Cox}(2001)}]{r:BC01}
\bibinfo{author}{\bibfnamefont{J.}~\bibnamefont{Berges}} \bibnamefont{and}
  \bibinfo{author}{\bibfnamefont{J.}~\bibnamefont{Cox}},
  \bibinfo{journal}{Phys. Lett.} \textbf{\bibinfo{volume}{B517}},
  \bibinfo{pages}{369} (\bibinfo{year}{2001}).

\bibitem[{\citenamefont{Aarts and Berges}(2001)}]{r:AB01}
\bibinfo{author}{\bibfnamefont{G.}~\bibnamefont{Aarts}} \bibnamefont{and}
  \bibinfo{author}{\bibfnamefont{J.}~\bibnamefont{Berges}},
  \bibinfo{journal}{Phys. Rev. D} \textbf{\bibinfo{volume}{64}},
  \bibinfo{pages}{105010} (\bibinfo{year}{2001}).

\bibitem[{\citenamefont{Berges}(2002)}]{r:B02}
\bibinfo{author}{\bibfnamefont{J.}~\bibnamefont{Berges}},
  \bibinfo{journal}{Nuc. Phys. A} \textbf{\bibinfo{volume}{699}},
  \bibinfo{pages}{847} (\bibinfo{year}{2002}).

\bibitem[{\citenamefont{Wilson}(1973)}]{r:W73}
\bibinfo{author}{\bibfnamefont{K.~G.} \bibnamefont{Wilson}},
  \bibinfo{journal}{Phys. Rev. D} \textbf{\bibinfo{volume}{7}},
  \bibinfo{pages}{2911} (\bibinfo{year}{1973}).

\bibitem[{\citenamefont{Coleman et~al.}(1974)\citenamefont{Coleman, Jackiw, and
  Politzer}}]{r:CJP}
\bibinfo{author}{\bibfnamefont{S.}~\bibnamefont{Coleman}},
  \bibinfo{author}{\bibfnamefont{R.}~\bibnamefont{Jackiw}}, \bibnamefont{and}
  \bibinfo{author}{\bibfnamefont{H.~D.} \bibnamefont{Politzer}},
  \bibinfo{journal}{Phys. Rev. D} \textbf{\bibinfo{volume}{10}},
  \bibinfo{pages}{2491} (\bibinfo{year}{1974}).

\bibitem[{\citenamefont{Schnitzer}(1974)}]{r:S74}
\bibinfo{author}{\bibfnamefont{H.~J.} \bibnamefont{Schnitzer}},
  \bibinfo{journal}{Phys. Rev. D} \textbf{\bibinfo{volume}{10}},
  \bibinfo{pages}{2042} (\bibinfo{year}{1974}).

\bibitem[{\citenamefont{Chang}(1975)}]{r:C75}
\bibinfo{author}{\bibfnamefont{S.-J.} \bibnamefont{Chang}},
  \bibinfo{journal}{Phys. Rev. D} \textbf{\bibinfo{volume}{12}},
  \bibinfo{pages}{1071} (\bibinfo{year}{1975}).

\bibitem[{\citenamefont{Cooper et~al.}(1986)\citenamefont{Cooper, Pi, and
  Stancioff}}]{r:CPS86}
\bibinfo{author}{\bibfnamefont{F.}~\bibnamefont{Cooper}},
  \bibinfo{author}{\bibfnamefont{S.~Y.} \bibnamefont{Pi}}, \bibnamefont{and}
  \bibinfo{author}{\bibfnamefont{P.~N.} \bibnamefont{Stancioff}},
  \bibinfo{journal}{Phys. Rev. D} \textbf{\bibinfo{volume}{34}},
  \bibinfo{pages}{3831} (\bibinfo{year}{1986}).

\bibitem[{\citenamefont{Cooper and Mottola}(1987)}]{r:CMprd36}
\bibinfo{author}{\bibfnamefont{F.}~\bibnamefont{Cooper}} \bibnamefont{and}
  \bibinfo{author}{\bibfnamefont{E.}~\bibnamefont{Mottola}},
  \bibinfo{journal}{Phys. Rev. D} \textbf{\bibinfo{volume}{36}},
  \bibinfo{pages}{3114} (\bibinfo{year}{1987}).

\bibitem[{\citenamefont{Pi and Samiullah}(1987)}]{r:PS86}
\bibinfo{author}{\bibfnamefont{S.~Y.} \bibnamefont{Pi}} \bibnamefont{and}
  \bibinfo{author}{\bibfnamefont{M.}~\bibnamefont{Samiullah}},
  \bibinfo{journal}{Phys. Rev. D} \textbf{\bibinfo{volume}{36}},
  \bibinfo{pages}{3128} (\bibinfo{year}{1987}).

\bibitem[{\citenamefont{Boyanovsky et~al.}(1994)\citenamefont{Boyanovsky,
  de~Vega, and Holman}}]{r:BVH94}
\bibinfo{author}{\bibfnamefont{D.}~\bibnamefont{Boyanovsky}},
  \bibinfo{author}{\bibfnamefont{H.~J.} \bibnamefont{de~Vega}},
  \bibnamefont{and} \bibinfo{author}{\bibfnamefont{R.}~\bibnamefont{Holman}},
  \bibinfo{journal}{Phys. Rev. D} \textbf{\bibinfo{volume}{49}},
  \bibinfo{pages}{2769} (\bibinfo{year}{1994}).

\bibitem[{\citenamefont{Blagoev et~al.}(2001)\citenamefont{Blagoev, Cooper,
  Dawson, and Mihaila}}]{r:BCDM01}
\bibinfo{author}{\bibfnamefont{K.~B.} \bibnamefont{Blagoev}},
  \bibinfo{author}{\bibfnamefont{F.}~\bibnamefont{Cooper}},
  \bibinfo{author}{\bibfnamefont{J.~F.} \bibnamefont{Dawson}},
  \bibnamefont{and} \bibinfo{author}{\bibfnamefont{B.}~\bibnamefont{Mihaila}},
  \bibinfo{journal}{Phys. Rev. D} \textbf{\bibinfo{volume}{64}},
  \bibinfo{pages}{125003} (\bibinfo{year}{2001}).

\bibitem[{\citenamefont{Mihaila et~al.}(2001)\citenamefont{Mihaila, Dawson, and
  Cooper}}]{r:MCD01}
\bibinfo{author}{\bibfnamefont{B.}~\bibnamefont{Mihaila}},
  \bibinfo{author}{\bibfnamefont{J.~F.} \bibnamefont{Dawson}},
  \bibnamefont{and} \bibinfo{author}{\bibfnamefont{F.}~\bibnamefont{Cooper}},
  \bibinfo{journal}{Phys. Rev. D} \textbf{\bibinfo{volume}{63}},
  \bibinfo{pages}{096003} (\bibinfo{year}{2001}).

\bibitem[{\citenamefont{Das and Kaku}(1978)}]{r:DasKaku78}
\bibinfo{author}{\bibfnamefont{A.}~\bibnamefont{Das}} \bibnamefont{and}
  \bibinfo{author}{\bibfnamefont{M.}~\bibnamefont{Kaku}},
  \bibinfo{journal}{Phys. Rev. D} \textbf{\bibinfo{volume}{18}},
  \bibinfo{pages}{4540} (\bibinfo{year}{1978}).

\bibitem[{\citenamefont{Moshe and Zinn-Justin}(2003)}]{ref:MMZJ}
\bibinfo{author}{\bibfnamefont{M.}~\bibnamefont{Moshe}} \bibnamefont{and}
  \bibinfo{author}{\bibfnamefont{J.}~\bibnamefont{Zinn-Justin}},
  \bibinfo{journal}{Nuc. Phys. B} \textbf{\bibinfo{volume}{648}},
  \bibinfo{pages}{131} (\bibinfo{year}{2003}).

\bibitem[{\citenamefont{Feinberg et~al.}(2005)\citenamefont{Feinberg, Moshe,
  and Smolkin}}]{ref:FMS}
\bibinfo{author}{\bibfnamefont{J.}~\bibnamefont{Feinberg}},
  \bibinfo{author}{\bibfnamefont{M.}~\bibnamefont{Moshe}}, \bibnamefont{and}
  \bibinfo{author}{\bibfnamefont{M.}~\bibnamefont{Smolkin}},
  \bibinfo{journal}{Inter. J. Mod. Phys. A} \textbf{\bibinfo{volume}{20}},
  \bibinfo{pages}{4475} (\bibinfo{year}{2005}).

\bibitem[{\citenamefont{Shifman et~al.}(1999)\citenamefont{Shifman, Vainshtein,
  and Voloshin}}]{ref:ShifVainVolo99}
\bibinfo{author}{\bibfnamefont{M.}~\bibnamefont{Shifman}},
  \bibinfo{author}{\bibfnamefont{A.}~\bibnamefont{Vainshtein}},
  \bibnamefont{and} \bibinfo{author}{\bibfnamefont{M.}~\bibnamefont{Voloshin}},
  \bibinfo{journal}{Phys. Rev. D} \textbf{\bibinfo{volume}{59}},
  \bibinfo{pages}{045016} (\bibinfo{year}{1999}).

\bibitem[{\citenamefont{Ivanov et~al.}(2005{\natexlab{a}})\citenamefont{Ivanov,
  Riek, , and Knoll}}]{r:IRK}
\bibinfo{author}{\bibfnamefont{Y.~B.} \bibnamefont{Ivanov}},
  \bibinfo{author}{\bibfnamefont{F.}~\bibnamefont{Riek}}, , \bibnamefont{and}
  \bibinfo{author}{\bibfnamefont{J.}~\bibnamefont{Knoll}},
  \bibinfo{journal}{Phys. Rev. D} \textbf{\bibinfo{volume}{71}},
  \bibinfo{pages}{105016} (\bibinfo{year}{2005}{\natexlab{a}}).

\bibitem[{\citenamefont{Ivanov et~al.}(2005{\natexlab{b}})\citenamefont{Ivanov,
  Riek, van Hees, and Knoll}}]{r:IRHK}
\bibinfo{author}{\bibfnamefont{Y.~B.} \bibnamefont{Ivanov}},
  \bibinfo{author}{\bibfnamefont{F.}~\bibnamefont{Riek}},
  \bibinfo{author}{\bibfnamefont{H.}~\bibnamefont{van Hees}}, \bibnamefont{and}
  \bibinfo{author}{\bibfnamefont{J.}~\bibnamefont{Knoll}},
  \bibinfo{journal}{Phys. Rev. D} \textbf{\bibinfo{volume}{72}},
  \bibinfo{pages}{036008} (\bibinfo{year}{2005}{\natexlab{b}}).

\end{thebibliography}
%
%
\end{document}